\documentclass[12pt]{article}
\usepackage{amsfonts}
\usepackage{amssymb}
\usepackage{amsmath}
\usepackage{color}
\usepackage{graphics}
\usepackage{graphicx}
\usepackage{figlatex}

\def\hybrid{\topmargin -20pt    \oddsidemargin 0pt
        \headheight 0pt \headsep 0pt
        \textwidth 6.25in       
        \textheight 9.25in       
        \marginparwidth .875in
        \parskip 5pt plus 1pt   \jot = 1.5ex}

\hybrid

\def\baselinestretch{1.2}

\catcode`\@=11

\def\marginnote#1{}
\newcount\hour
\newcount\minute
\newtoks\amorpm
\hour=\time\divide\hour by60
\minute=\time{\multiply\hour by60 \global\advance\minute by-\hour}
\edef\standardtime{{\ifnum\hour<12 \global\amorpm={am}%
        \else\global\amorpm={pm}\advance\hour by-12 \fi
        \ifnum\hour=0 \hour=12 \fi
        \number\hour:\ifnum\minute<10 0\fi\number\minute\the\amorpm}}
\edef\militarytime{\number\hour:\ifnum\minute<10 0\fi\number\minute}

\def\draftlabel#1{{\@bsphack\if@filesw {\let\thepage\relax
   \xdef\@gtempa{\write\@auxout{\string
      \newlabel{#1}{{\@currentlabel}{\thepage}}}}}\@gtempa
   \if@nobreak \ifvmode\nobreak\fi\fi\fi\@esphack}
        \gdef\@eqnlabel{#1}}
\def\@eqnlabel{}
\def\@vacuum{}
\def\draftmarginnote#1{\marginpar{\raggedright\scriptsize\tt#1}}

\def\draft{\oddsidemargin -.5truein
        \def\@oddfoot{\sl preliminary draft \hfil
        \rm\thepage\hfil\sl\today\quad\militarytime}
        \let\@evenfoot\@oddfoot \overfullrule 3pt
        \let\label=\draftlabel
        \let\marginnote=\draftmarginnote
   \def\@eqnnum{(\theequation)\rlap{\kern\marginparsep\tt\@eqnlabel}%
\global\let\@eqnlabel\@vacuum}  }

\def\preprint{\twocolumn\sloppy\flushbottom\parindent 2em
        \leftmargini 2em\leftmarginv .5em\leftmarginvi .5em
        \oddsidemargin -.5in    \evensidemargin -.5in
        \columnsep .4in \footheight 0pt
        \textwidth 10.in        \topmargin  -.4in
        \headheight 12pt \topskip .4in
        \textheight 6.9in \footskip 0pt
        \def\@oddhead{\thepage\hfil\addtocounter{page}{1}\thepage}
        \let\@evenhead\@oddhead \def\@oddfoot{} \def\@evenfoot{} }

\def\numberbysection{\@addtoreset{equation}{section}
        \def\theequation{\thesection.\arabic{equation}}}

\def\underline#1{\relax\ifmmode\@@underline#1\else
        $\@@underline{\hbox{#1}}$\relax\fi}

\def\titlepage{\@restonecolfalse\if@twocolumn\@restonecoltrue\onecolumn
     \else \newpage \fi \thispagestyle{empty}\c@page\z@
        \def\thefootnote{\fnsymbol{footnote}} }

\def\endtitlepage{\if@restonecol\twocolumn \else \newpage \fi
        \def\thefootnote{\arabic{footnote}}
        \setcounter{footnote}{0}}  

\catcode`@=12
\relax

\def\figcap{\section*{Figure Captions\markboth
        {FIGURECAPTIONS}{FIGURECAPTIONS}}\list
        {Figure \arabic{enumi}:\hfill}{\settowidth\labelwidth{Figure
999:}
        \leftmargin\labelwidth
        \advance\leftmargin\labelsep\usecounter{enumi}}}
 \relax
\def\tablecap{\section*{Table Captions\markboth
        {TABLECAPTIONS}{TABLECAPTIONS}}\list
        {Table \arabic{enumi}:\hfill}{\settowidth\labelwidth{Table
999:}
        \leftmargin\labelwidth
        \advance\leftmargin\labelsep\usecounter{enumi}}}
 \relax
\def\reflist{\section*{References\markboth
        {REFLIST}{REFLIST}}\list
        {[\arabic{enumi}]\hfill}{\settowidth\labelwidth{[999]}
        \leftmargin\labelwidth
        \advance\leftmargin\labelsep\usecounter{enumi}}}
 \relax
\makeatletter
\newcounter{pubctr}
\def\publist{\@ifnextchar[{\@publist}{\@@publist}}
\def\@publist[#1]{\list
        {[\arabic{pubctr}]\hfill}{\settowidth\labelwidth{[999]}
        \leftmargin\labelwidth
        \advance\leftmargin\labelsep
        \@nmbrlisttrue\def\@listctr{pubctr}
        \setcounter{pubctr}{#1}\addtocounter{pubctr}{-1}}}
\def\@@publist{\list
        {[\arabic{pubctr}]\hfill}{\settowidth\labelwidth{[999]}
        \leftmargin\labelwidth
        \advance\leftmargin\labelsep
        \@nmbrlisttrue\def\@listctr{pubctr}}}
 \relax
\makeatother
\newskip\humongous \humongous=0pt plus 1000pt minus 1000pt

\newif\ifdtup

\relax

\def\be{\begin{equation}}
\def\ee{\end{equation}}
\def\ba{\begin{eqnarray}}
\def\ea{\end{eqnarray}}

\def\del{\partial}

\def\r{\rho}
\def\a{\alpha}

\def\d{\delta}

\def\e{\epsilon}

\def\th{\theta}

\def\m{\mu}

\def\s{\sigma}
\def\S{\Sigma}

\def\no{\noindent}

\def\qq{\qquad}

\def\IR{\relax{\rm I\kern-.18em R}}

\def \ha {{1\over 2}}

\def \ov {\over}

\def\II{\relax{\rm 1\kern-.35em1}}
\def\IR{\relax{\rm I\kern-.18em R}}
\def\inv{^{\raise.15ex\hbox{${\scriptscriptstyle -}$}\kern-.05em 1}}

\begin{document}
\renewcommand{\theequation}{\arabic{equation}}

\renewcommand{\theequation}{\thesection.\arabic{equation}}
\csname @addtoreset\endcsname{equation}{section}

\newcommand{\beq}{\begin{equation}}
\newcommand{\eeq}[1]{\label{#1}\end{equation}}
\newcommand{\ber}{\begin{eqnarray}}
\newcommand{\eer}[1]{\label{#1}\end{eqnarray}}
\newcommand{\eqn}[1]{(\ref{#1})}
\begin{titlepage}
\begin{center}

\hfill CERN-PH-TH 2008/061
\vskip  0.5in

{\large \bf  Supersymmetric moduli of the  $SU(2) \times \mathbb{R}_\phi$ \\
linear dilaton background and NS5-branes}

\vskip 0.4in

{\bf Nikolaos Prezas$^{1a}$} and
{\bf Konstadinos Sfetsos}$^{1,2b}$ \vskip 0.1in

${}^1\!$
Theory Unit, Physics Department, CERN\\
1211 Geneva, Switzerland\\

\vskip .2in

${}^2\!$
Department of Engineering Sciences, University of Patras\\
26110 Patras, Greece\\

\end{center}

\vskip .3in

\centerline{\bf Synopsis}

\no
We study several classes of marginal deformations of the conformal field theory
$SU(2)_k \times  \mathbb{R}_\phi$. This theory describes
the near-horizon region of a stack of parallel and coincident NS5-branes and is related
holographically to  little string theory. We investigate the supersymmetry
properties of these deformations
and we elucidate their r\^ole in the context of holography.
The conformal field theory moduli space contains ``non-holographic" 
operators that do not seem
to have a simple interpretation in little string theory.
Subsequently, we analyze several NS5-brane configurations in terms of
$SU(2)_k \times  \mathbb{R}_\phi$ deformations. We discuss in detail
  interesting phenomena,
like the excision of
the strong coupling region associated with the linear dilaton  and the
manifestation of the symmetries of an NS5-brane setup in the deforming operators.
Finally, we present a class of conformally hyperk\"ahler geometries 
that arise as ``non-holographic" deformations of $SU(2)_k \times  \mathbb{R}_\phi$.

\no

\no

\vfill
\begin{flushleft}
 {\footnotesize
 $^a$nikolaos.prezas@cern.ch,
 $^b$sfetsos@upatras.gr}
 \end{flushleft}
\end{titlepage}
\vfill
\newpage
\setcounter{footnote}{0}
\renewcommand{\thefootnote}{\arabic{footnote}}


\setcounter{section}{0}

\def\baselinestretch{1.2}
\baselineskip 20 pt
\noindent


\tableofcontents

\section{\boldmath Introduction \unboldmath}

String theory backgrounds that admit an exact conformal field theory (CFT)
description are of particular interest since their physical properties can be
analyzed to all orders in $\alpha'$. The situation is even more interesting when
these backgrounds are created by the
backreaction of a configuration of branes.
In this case, deformations of the CFT correspond to deformations of the brane
system. It often happens that some of
the latter can be visualized as changes in the geometry of the branes,
thereby leading to a
very intuitive geometrical picture of the CFT moduli space.

\no
Unfortunately, the number of brane systems that admit an exact CFT description
is rather small. First of all, configurations with D-branes source Ramond--Ramond
fields and, as is well-known, it is notoriously difficult to obtain a useful CFT description of such backgrounds.
However, even when solely NS5-branes are present, there are so far only two
instances where the underlying CFT is known explicitly. The first is a configuration
of $k$ parallel NS5-branes put at the same point in their transverse $\mathbb{R}^4$ space.
The near-horizon region of this system is described by the
Callan--Harvey--Strominger (CHS) theory $SU(2)_k \times \mathbb{R}_\phi$
\cite{Callan:1991dj}.  We will provide a brief reminder on this theory in the next section.
The second instance is that of $k$ parallel NS5-branes put uniformly on a circle in  $\mathbb{R}^4$.
In this case, their near-horizon region is described by the coset CFT
$SU(2)_k /U(1) \times SL(2)_k / U(1)$ orbifolded by $\mathbb{Z}_k$
\cite{Sfetsos:1998xd}.
Both of these theories support the ${\cal N}=4$ superconformal  algebra,
since the associated NS5-brane systems are 1/2 BPS (they preserve 16
supercharges in type II string theories and 8 supercharges in the heterotic string).

\no
Geometric deformations of the system of NS5-branes away from
the point or the circle distribution are associated with exactly marginal operators
in the underlying CFT. The reason is that moving the NS5-branes away from their
original locations yields a configuration that is also a solution of the equations of motion, continuously connected
to the original one,  and the space
of such solutions is generically identified with the space of exactly marginal deformations
of the CFT. Furthermore, since an arbitrary configuration of NS5-branes
in $\mathbb{R}^4$ preserves the same amount of supersymmetry as the point-like
configuration, we are naturally lead to consider only the  deformations of the CHS theory
that preserve all of the original ${\cal N}=4$ worldsheet supersymmetry.

\no
The connection between deformations of the NS5-brane system and marginal
operators in the CFT can be established either directly or through the
use of holography. The first approach is based on the fact that changes of the
original locations of the NS5-branes correspond to perturbations of the original
supergravity background that subsequently induce
deformations of the associated worldsheet $\sigma$ model. One can
read these $\sigma$ model deformations and express them in terms of operators
of the original undeformed theory. The last step is performed by employing the semiclassical expression
of these operators in terms of $\sigma$ model target space fields. This approach
to NS5-brane deformations
was initiated in \cite{Marios Petropoulos:2005wu}, where the operators
in $SU(2)_k /U(1) \times SL(2,\mathbb{R})_k / U(1)$
that trigger an elliptical perturbation of the circular NS5-brane system
were uncovered.

\no
The second approach is based on the fact that the decoupled worldvolume
theory on the NS5-branes, known as little string theory (LST), admits
a holographic description  in terms of
string theory on the near-horizon limit of the background
generated by the NS5-branes \cite{Aharony:1998ub}. The conjectured holography
implies  a correspondence between operators in LST
and vertex operators in the dual string theory background \cite{Aharony:2003vk,
Aharony:2004xn}. Since the moduli
space of the geometric NS5-brane deformations is the moduli space of LST and the latter is parametrized
by gauge invariant LST operators, we see that one can associate in this way
deformations of the NS5-branes with operators in the underlying CFT. At the level
of holography this association is done by using the symmetries of the two
sides of the correspondence.  However, symmetry matchings do not constitute a proof
and one would like to substantiate the holographic dictionary between operators
in a more explicit way.
This was achieved in \cite{Fotopoulos:2007rm}
where the first approach, based on the $\sigma$ model description of the deformed
NS5-brane background, was used to validate the holographic correspondence
in the semiclassical limit of large $k$.

\no
In this paper we investigate some generic issues pertaining
to deformations of the CHS theory and in conjunction with its NS5-brane interpretation.
First we will perform a study of the supersymmetry properties of several types
of marginal operators of the CHS. We will be particularly interested
in operators that preserve the original ${\cal N}=4$  superconformal symmetry
of the CHS model, as these operators can in  principle correspond to geometric
deformations of the NS5-branes. Surprisingly, we will also uncover some other
classes of supersymmetry preserving marginal deformations that do not seem to have
holographic counterparts. Although we will elucidate the physical effects of some
of them with
some simple examples, presented in the last section,
the precise understanding of their  interpretation in terms of
NS5-branes and  LST is left for future work.
Subsequently, we will consider a set of  NS5-brane configurations
 that arise as deformations of the point-like setup and, therefore, can be described by
 marginal operators  in the CHS theory. Our objective would be to show
 how the physical and geometrical properties of these configurations
 are encoded in the corresponding CFT operators. This analysis will illuminate further
 the fascinating interplay between spacetime and CFT physics.

\section{\boldmath Supersymmetric operators in the CHS background\unboldmath}

In this section we study the supersymmetry properties of a class of marginal operators of the CHS background. First, we find conditions  for these operators to be  chiral or antichiral
primaries and subsequently we check which of these operators yield supersymmetry
preserving deformations. Since extended worldsheet supersymmetry is
necessary for spacetime (i.e.~worldvolume) supersymmetry, we check first the former and then
perform a test of the latter. In the course
of our analysis we will uncover that some operators that do not seem to have a
holographic intepretation preserve also maximal supersymmetry.

\subsection{Generalities}
The holographic description of LST is based on the correspondence
between BPS operators in LST and vertex operators in the CHS background
$SU(2)_k \times \mathbb{R}_\phi$ \cite{Aharony:2003vk,Aharony:2004xn}.
The  latter contains a linear dilaton along the $\phi$ direction with
background charge $q=\sqrt{\frac{2}{k}}$, where $k$ is the number
of NS5-branes, and a ${\cal N}=1$
supersymmetric $SU(2)$ WZW model at level $k$ generated by
affine currents ${\cal J}^a, a=1,2,3$.
A class of BPS operators in LST consists of
$\widetilde {\rm  tr} (X^{i_1}X^{i_2}\cdots X^{i_{2j+2}})$
with $j=0,\frac{1}{2},1,\ldots,\frac{k-2}{2}$ and where
$X^i, i=6,7,8,9$ are  scalar fields in the adjoint representation of $SU(k)$
whose eigenvalues parametrize
the transverse positions of the NS5-branes. In order that the LST operators
are in a short multiplet of spacetime supersymmetry, only the
traceless and symmetric components in the indices $i_1,\ldots,i_{2j+2}$ should be kept.
The tilde on the trace means that we should not consider the standard single trace
but its combination with multi-traces. This subtlety, however, will not play any
r\^ole in the considerations of this section.

\no
 The dictionary proposed in
 \cite{Aharony:2003vk, Aharony:2004xn} and tested in a non-trivial
 setup in \cite{Fotopoulos:2007rm}
 states the correspondence
  \begin{equation}\label{holodic}
\widetilde {\rm  tr} (X^{i_1}X^{i_2}\cdots X^{i_{2j+2}}) \longleftrightarrow
  (\psi\bar \psi \Phi_j)_{j+1;m,\bar m}
e^{-q a_j \phi}\ ,
\end{equation}
where the right-hand side is an  operator in the CHS
theory.
The coefficient $a_j$ of the linear dilaton  vertex operator at the right
must be either $a_j=j+1$ or $a_j=-j$ in order that the actual deforming operators,
which arise from the action of the ${\cal N}=1$ supercharges on the 
operators at the right-hand side of (\ref{holodic}), are marginal
(see formula (\ref{lagdef}) below).
In the first case the operator
is normalizable\footnote{Notice that strictly speaking one cannot talk meaningfully
about normalizable operators in the CHS theory since they are supported 
in the strong coupling region $\phi \rightarrow -\infty$. However,
there is a 1-1 correspondence between operators of the $SU(2)_k \times \mathbb{R}_\phi$
theory and of  the non-singular coset CFT $SL(2)_k/U(1) \times SU(2)_k/U(1)$,
so that all of our subsequent discussion can be trivially generalized to
the physically more reliable coset theory.}
and hence it corresponds to a situation where the
dual  LST operator acquires a vacuum expectation value. 

\no
The way one
associates geometric deformations of the NS5-branes to CHS operators using the
above holographic correspondence is the following.
The original configuration of NS5-branes put at the point $x^6=x^7=x^8=x^9=0$
is described, in the near-horizon limit, by the unperturbed CHS theory.
A generic point in the moduli space,
which corresponds to separating
the branes in
their transverse $\mathbb{R}^4$, thereby turning on non-vanishing expectation
values for the scalars $X^i$, is described by a deformation of the original CFT with
operators that can be found using the correspondence (\ref{holodic})
(according to formula (\ref{lagdef}) below). Notice that we consider only deformations
that leave invariant the center of mass of the NS5-brane system, in other words
we always assume that
${\rm tr} (X^i)=0$. The reason is that the associated $U(1)$ degree of freedom
in LST is frozen and decouples, so that there is no normalizable mode corresponding
to it. The other value of $a_j$ that yields also a marginal deformation,
 i.e.~$a_j=-j$, corresponds to a
non-normalizable deformation of the CHS theory that triggers
a perturbation of the LST with the operator at the left-hand side.

\no
In order to write the CFT operators explicitly, we
decompose the supersymmetric WZW model into a bosonic
$SU(2)_{k-2}$ WZW model at level $k-2$, whose affine currents we will denote by $J^i$, and
three free fermions $\psi^a, \; a=1,2,3$ in the adjoint of $SU(2)$.
Consequently, the  ${\cal N}=1$ affine currents can be written
as ${\cal J}^a=J^a- \frac{i}{2} \epsilon^{abc} \psi^b \psi^c$.
The field $\Phi_j$ is in general  a Virasoro primary of the
 bosonic $SU(2)_{k-2}$ WZW model and the notation
 $(\psi\bar \psi \Phi_j)_{j+1;m,\bar m}$ means that we should tensor
 the fermions $\psi_a$ to
  the bosonic primary  $\Phi_j$ into a primary of total spin $j+1$ and
$({\cal J}^3,\bar {\cal J}^3)=(m,\bar m)$.

\no
It will be practical to introduce
the complex fermiom combinations
$\psi^\pm=\frac{1}{\sqrt{2}}(\psi_1\pm i \psi_2)$ and also perform  the usual
change of basis for the $SU(2)_{k-2}$ currents $J^{\pm}=J^1\pm i
J^2$. The super-affine currents then read ${\cal J}^3=J^3 + \psi^+ \psi^-$ and $
{\cal J}^\pm = J^\pm \pm \sqrt{2} \psi^3 \psi^\pm$. Finally, we will use extensively
the $SU(2)_{k-2}$ current algebra at level $k-2$
\begin{equation}
\label{OPEJ}
\begin{array}{rcl}
\displaystyle{J^3(z)J^3(w)}& \sim &\displaystyle{\frac{k-2}{2 (z-w)^2}}
\ , \\\ \displaystyle{J^3(z)J^\pm(w)}&\sim &\displaystyle{\pm
\frac{J^\pm(w)}{z-w}} \ , \\\
\displaystyle{J^+(z)J^-(w)}&\sim &\displaystyle{\frac{k-2}{(z-w)^2}+\frac{2J^3(w)}{z-w}}
\ ,
\end{array}
\end{equation}
and the
action of the $SU(2)_{k-2}$ currents on the Virasoro primaries $\Phi_{j;m}$:
\ba
 J^3(z) \Phi_{j;m}(w) &\sim&  {m\ov z-w} \Phi_{j;m}(w)\ ,
\nonumber
\\
 J^\pm(z) \Phi_{j;m}(w) & \sim &   {j\mp m\ov z-w} \Phi_{j;m\pm 1}(w)\ .
\label{sdjop}
\ea

\no
Now, we can write explicitly
\begin{equation}
(\psi \bar \psi \Phi_j)_{j+1;m,\bar m} = N_j \bar N_j
\sum_{r,s=-1}^{1}
c_r(j,m) c_s(j,\bar m) \psi^r \bar \psi^s \Phi_{j;m-r,\bar m-s}\label{cleb}\ ,
\end{equation}
where we use the notation
$(\psi^1,\psi^0,\psi^{-1})\equiv(\psi^+,\psi^3,\psi^-)$
and the Clebsch--Gordan coefficients $c_r(j,m)$ are given by
\begin{eqnarray}
c_1(j,m) &=& - \frac{1}{\sqrt{2}}(j+m)(j+m+1)\ ,
\nonumber\\
c_0(j,m) &=&  (j+m+1)(j-m+1)\ ,
\label{hglq}\\
c_{-1}(j,m) &=& \frac{1}{\sqrt{2}}(j-m)(j-m+1)\ .
\nonumber
\end{eqnarray}
The Clebsch--Gordan coefficients are determined in terms
of the coefficients in the action of $J^\pm$ on the primaries $\Phi_{j;m}$. In our
case they differ from the more familiar form involving square roots
due to our conventions in
(\ref{sdjop}).
We have also introduced a convenient $j$-dependent normalization factor
given by
\begin{equation}
N_j = \bar N_j= \frac{1}{(2j+1)(2j+2)}\ .
\end{equation}

\subsection{Chiral and antichiral primaries}

\no
At this stage one could ask if the CFT operators in
 (\ref{holodic})
have any special properties. Since the CHS background exhibits ${\cal N}=4$
superconformal invariance,
a natural question is if there are any chiral or antichiral primaries among them.
Let us choose the ${\cal N}=2$ subalgebra generated by
the energy-momentum tensor
\begin{equation}\label{Tgen}
T=-\frac{1}{2}(\partial\phi)^2-\frac{1}{2} q
\partial^2\phi+\frac{J^i J^i}{k}-
\frac{1}{2}\psi^*\partial\psi-\frac{1}{2}\psi\partial\psi^*-\frac{1}{2}\psi^+\partial\psi^-
-\frac{1}{2}\psi^-\partial\psi^+\ ,
\end{equation}
the supercurrents
\begin{eqnarray}\label{CHSGpm2}
 G^+&=&i\psi\left(\partial\phi- q J_3- q \psi^+ \psi^- \right)+i q \partial\psi+ q J^-\psi^+\ ,
\nonumber\\
 G^-&=&i\psi^*\left(\partial\phi+ q J_3 + q \psi^+ \psi^- \right)+i q \partial\psi^*+ q J^+\psi^-\ ,
 \end{eqnarray}
 and the $U(1)$ R-current
 \begin{equation}
J_R = \psi \psi^* +  \psi^+ \psi^- =  -i \psi_\phi \psi_3 + \psi^+ \psi^- \ .\label{rcur}
\end{equation}
The fermion combinations
\begin{equation}\label{psidef}
\psi^\pm=\frac{1}{\sqrt{2}}(\psi_1\pm i \psi_2)\ , \qq
\psi=\frac{1}{\sqrt{2}}(\psi_\phi+i \psi_3) \ ,
\end{equation}
with $\psi_\phi$ being the superpartner of $\phi$, satisfy the following
operator product expansions
\begin{equation} \label{OPEpsi}
\psi(z)\psi^*(w)=\psi^+(z)\psi^-(w) \sim \frac{1}{z-w}\ .
\end{equation}
For considerations of spacetime supersymmetry, it will be also useful
to bosonize the above fermions as
\begin{equation}
\psi^\pm = e^{\pm i H_1}, \quad \psi= e^{i H_2}
\end{equation}
with $H_1$ and $H_2$ being canonically normalized bosons with OPEs
\begin{equation}
H_1(z) H_1(w) = H_2(z) H_2(w) \sim -\ln(z-w)\ , \quad H_1(z) H_2(w) = 0\ .
\end{equation}
Then the R-current can be written as
\begin{equation}
J_R = i \partial H_1 + i \partial H_2
\end{equation}
and (half of) the spacetime supercharges, which live on the 5+1-dimensional worldvolume
of the NS5-branes, are given by
\begin{equation}
Q^{\pm}_\alpha = \frac{1}{2 \pi i} \oint dz e^{-\frac{\varphi}{2}\pm \frac{i}{2}(H_1+H_2)}
{\cal S}_\alpha\ .\label{spacesusy}
\end{equation}
In this formula $\varphi$ stands for the bosonized superconformal ghosts
and ${\cal S}_\alpha$ are worldvolume spin fields in the ${\bf 4}$ of
$SO(5,1)$ whose explicit form
will not be necessary. For NS5-branes in type II theories
a similar set of spacetime supercharges arises from
the antiholomorphic sector, so that all together we have 16 spacetime
supersymmetries. Notice that in general
we will focus only on the holomorphic sector, since exactly the same expressions
hold for the antiholomorphic one, and from now on
we will suppress in most formulas all antiholomorphic indices
to avoid cluttering. 

\no
Recall that a field $\chi $ is primary of the ${\cal N}=2$ superconformal algebra
if it satisfies
\begin{eqnarray}
T(z)  \chi (w) &\sim & \frac{h}{(z-w)^2} \chi(w)+\frac{\partial \chi(w)}{z-w}\ ,
\nonumber\\
J_R(w) \chi(w) &\sim & \frac{q}{z-w} \chi(w)\ , \\
G^{\pm}(z) \chi(w) &\sim & \frac{1}{z-w} \widetilde \chi^{\pm}(w)\ ,
\nonumber
\end{eqnarray}
with $h$ being its conformal weight and $q$ its U(1) R-charge.
In addition, it is chiral (antichiral)  if its OPE with the supercurrent $G^+(z)$ $\left(G^-(z)\right)$
is regular \cite{LVW}. When a field is both chiral (antichiral) and primary
its conformal dimension is fixed in terms of its U(1)
R-charge as $h= |q|/2$. As a consequence $h$ is not renormalized as long as
superconformal invariance remains unbroken. In particular, chiral and antichiral primary operators
with $|q|=1$ which yield marginal deformations when acted with the ${\cal N}=1$
supercharge $G=\frac{1}{\sqrt{2}}(G^++G^-)$, actually give rise to
{\em exactly marginal} deformations.

\no
It is a straightforward exercise to check that
$(\psi\Phi_j)_{j+1;m}
e^{-q a_j \phi}$
is a superconformal primary when $m=j+1$ or
$m=-j-1$. Then, the corresponding operators
take the form $\psi^+ \Phi_{j;j} e^{-q \a_j \phi}$
and  $\psi^- \Phi_{j;-j} e^{-q \a_j \phi}$, respectively.  Notice that
these superconformal primaries are built on affine primaries
$\Phi_{j,\pm j}$ of the $SU(2)_{k-2}$ WZW model. Furthermore, out
of the class of operators $(\psi\Phi_j)_{j+1;m}
e^{-q a_j \phi}$,
only 
$\psi^+ \Phi_{j;m-1} e^{-q a_j \phi}$ and $\psi^- \Phi_{j;m+1} e^{-q a_j \phi}$
can have special chirality properties.
These operators are chiral (antichiral) when
 $m=a_j$  ($m=-a_j$). The final conclusion is that we have
 a set of chiral primaries given by  $\psi^+ \Phi_{j;j} e^{-q (j+1) \phi}$ along with their
conjugates  $\psi^- \Phi_{j;-j} e^{-q (j+1) \phi}$ which are antichiral primaries.
It is interesting to note  that
only normalizable operators, i.e. with $a_j=j+1$, can be chiral or antichiral primaries.

\no
Another operator we could consider is $\psi \Phi_{j;m} e^{-q a_j \phi}$ and its conjugate,
although only their real part $\psi_3 \Phi_{j;m} e^{-q a_j \phi}$
appears in the
holographic dictionary. These operators are primary when $m=1-a_j$ and they
are chiral (antichiral) when $m=-j$ ($m=j$). Hence we conclude
that $\psi \Phi_{j;-j} e^{-q (j+1)} \phi$  $\left (\psi^* \Phi_{j;j} e^{-q (j+1)}
\right)$ is a chiral (antichiral) primary.  Note that their non-normalizable counterparts
are not chiral or antichiral primaries.

\no
The fact that the non-normalizable versions of the chiral (antichiral)
primaries are not also chiral (antichiral) primaries seems a bit
puzzling at first sight.
For instance,  although  $\psi^+ \Phi_{j;j} e^{-q (j+1)\phi}$ and
$\psi \Phi_{j;-j} e^{-q (j+1) \phi}$ are chiral primary (and their conjugates
antichiral primary), their non-normalizable versions, that share the same $h$
and $q$, are not. This seems to violate the standard argument that an operator with
$h=q/2$ ($h=-q/2$) is chiral (antichiral) primary. This argument is based on the observation that
\cite{LVW}
\begin{equation}
\langle \chi | \{ G^+_{-\frac{1}{2}}, G^-_{\frac{1}{2}}\}|\chi\rangle = \langle \chi | 2 L_0 - (J_{R})_{0} |\chi\rangle = (2h-q) \langle \chi |\chi\rangle\ ,
\label{chiraa}
\end{equation}
where the
${\cal N}=2$ superconformal algebra was used. If $h=q/2$ one gets
$\langle \chi | \{ G^+_{-\frac{1}{2}}, G^-_{\frac{1}{2}}\}|\chi\rangle = 0$
and using hermiticity of the supercurrents
$(G^\pm_r)^\dagger=G^{\mp}_{-r}$ along with positivity of the inner product
leads to $G^+_{-\frac{1}{2}} |\chi\rangle = G^-_{\frac{1}{2}} |\chi \rangle =0$.
The resolution of the puzzle is that the linear dilaton CFT contains non-unitary representations
that correspond to fields  with
negative conformal weights.

\no
For instance,  $\psi^+ \Phi_{j;j} e^{q j\phi}$ is non-chiral and therefore,
if $|\chi\rangle$ is the corresponding state, we have
$G^+_{-\frac{1}{2}} |\chi\rangle \neq 0$.
Indeed, we obtain
\begin{equation}
|\hat \chi \rangle= G^+_{-\frac{1}{2}} |\chi\rangle = -i q (2j+1) \psi \psi^+ \Phi_{j;j}
e^{q j \phi} |\Omega\rangle\ ,
\end{equation}
where $|\Omega\rangle$ is the vacuum. This state,
however, satisfies $G^-_{\frac{1}{2}} |\hat \chi \rangle =0$ since there is no
second order pole between $G^-(z)$ and the operator $\psi \psi^+ \Phi_{j;j}
e^{q j \phi} (w)$. Hence, although the operator $\psi^+ \Phi_{j;j} e^{q j\phi}$  is non-chiral, it
still has $h=q/2=1/2$ and (\ref{chiraa}) is obeyed.
Similarly, $\psi \Phi_{j;-j} e^{q j \phi}$ is chiral but not primary since the state
$|\zeta \rangle$ it creates is not annihilated by $G^-_{\frac{1}{2}}$:
\begin{equation}
|\hat \zeta \rangle = G^-_{\frac{1}{2}} |\zeta \rangle = -i q (2j+1) \Phi_{j;-j} e^{q j \phi}|\Omega\rangle\ .
\end{equation}
However, we can again  check that
$G^+_{-\frac{1}{2}} |\hat \zeta \rangle =0$ since the OPE of $G^+(z)$ with
$\Phi_{j;-j} e^{q j \phi}$ is regular. The existence of non-unitary representations
of the linear dilaton CFT
underlies both effects as it is obvious from the fact that the
zero-norm states $|\hat \chi \rangle$ and $|\hat \zeta \rangle$
vanish when the  background charge $q$ is zero.

\subsection{Supersymmetric deformations}

\no
According to the prescription given in \cite{Aharony:2003vk,Aharony:2004xn},
when the operators  $ \widetilde {\rm tr} (X^{i_1}X^{i_2}\cdots X^{i_{2j+2}})$ obtain
non-zero VEVs  the original  Lagrangian ${\cal L}_0$ of the holographically
dual  conformal field theory is perturbed to
\begin{equation}
{\cal L}={\cal L}_0+\sum_{j=0}^{{\frac{k-2}{2}}}\sum_{m,\bar m=-(j+1)}^{j+1}
\Big(\lambda_{j;m,\bar m} G_{-\frac{1}{2}}\bar G_{-\frac{1}{2}}
 (\psi\bar \psi \Phi_j^{{\rm su}})_{j+1;m,\bar m}
e^{-q(j+1)\phi} +{\rm c.c.} \Big)\label{lagdef}\ .
\end{equation}
Here $G(z)$ is the ${\cal N}=1$ supercurrent
 \begin{equation}
G= i \psi_\phi \partial\phi + q \psi_3 J_3 + q \psi_3 \psi^+\psi^-+i q \partial\psi_\phi+
\frac{q}{\sqrt{2}}(J^-\psi^++J^+\psi^-)\ ,\label{n1sc}
\end{equation}
and the couplings $\lambda_{j;m,\bar m}$ are specified in terms of
  $ \widetilde {\rm tr} (X^{i_1}X^{i_2}\cdots X^{i_{2j+2}})$ in a way
  that we will make precise in the next section.
  Notice that
  by construction the deformation preserves ${\cal N}=(1,1)$ superconformal
  invariance.

\no
The action of $G_{-\frac{1}{2}}$ can be read from the simple pole of
$G(z)$ in its OPE with $(\psi\Phi_j)_{j+1;m}
e^{-q a_j \phi}$
and it yields a piece without fermions
and a piece bilinear in the fermions.
The first piece reads
\begin{equation}
q N_j \sum_{r=-1}^1 c_r(j,m) \lambda_r J^r \Phi_{j;m-r}
e^{-a_j q \phi}\ ,
\label{bosonic}
\end{equation}
where  $(J^{\pm 1}, J^0)\equiv(J^\pm,J^3)$ and $\lambda_0=1, \lambda_{\pm 1} =\frac{1}{
\sqrt{2}}$.
The fermion bilinear term is
\begin{equation}
q N_j   \Bigg(   \Big( \sum_{r=-1}^1 i a_j c_r \psi_\phi
\psi^r  \Phi_{j;m-r} \Big)+ d_1 \psi_3 \psi^+  \Phi_{j;m-1}
+ d_{-1} \psi_3 \psi^-  \Phi_{j;m+1}+d_0 \psi^+ \psi^-  \Phi_{j;m}
\Bigg) e^{-a_j q \phi}\ ,
\label{bilinear}
\end{equation}
where we defined the combinations
\begin{eqnarray}
d_{\pm 1}&=& m c_{\pm 1} - \frac{c_0}{\sqrt{2}} (j\pm m)\ ,
\nonumber\\
d_0&=& c_0 + \frac{1}{\sqrt{2}}\Big(c_{-1} (j+m+1)-c_{1} (j-m+1)\Big)\ .
\label{n1con2}
\end{eqnarray}
Notice that one could reverse the logic and start with an ansatz for the deformation
that is the sum of (\ref{bosonic}) and (\ref{bilinear}) with arbitrary coefficients
$c_r$ and $d_r$. Then, the equations (\ref{n1con2}) could be thought
of as conditions for preserving ${\cal N}=1$ worldsheet supersymmetry.

\subsubsection{${\cal N}=2$ supersymmetry}

\no
We will first uncover the conditions for  the sum of  the deformations  (\ref{bosonic})
and (\ref{bilinear}) to
preserve
 ${\cal N}=2$ supersymmetry and then
extend the analysis to ${\cal N}=4$. If  ${\cal N}=2$ is preserved, the deformation
should be annihilated by both supercharges $G^\pm_{-\frac{1}{2}}$
or, equivalently, by $G_{-\frac{1}{2}}$ and $G^3_{-\frac{1}{2}}$
where
\begin{equation}
G^3=-\frac{i}{\sqrt{2}} (G^+-G^-) \ .
\end{equation}
Since the deformation we consider arises from the action of  $G_{-\frac{1}{2}}$
on $(\psi\Phi_j)_{j+1;m}
e^{-q a_j \phi}$, it is automatically annihilated by $G_{-\frac{1}{2}}$.
Furthermore, a sufficient condition for the deformation
being annihilated by $G^3_{-\frac{1}{2}}$
is that it has zero R-charge,
as can be seen by the following ${\cal N}=2$
commutation relation
\begin{equation}
[ (J_R)_0,G_{-\frac{1}{2}}] = i G^3_{-\frac{1}{2}}\ .
\end{equation}
Actually, if we were only interested in preserving ${\cal N}=2$ supersymmetry, 
it would be enough to just demand definite R-charge. However, we want to preserve
${\cal N}=2$ superconformal invariance and since the R-symmetry is part of
the ${\cal N}=2$ SCFT algebra, the deformations we consider have to be neutral.
Notice also that the condition we just formulated 
is not necessary and, in principle, it could miss
some supersymmetric deformations. However,
we will soon establish that for the operators under consideration it is actually necessary, besides
being sufficient.

\no
The purely bosonic part (\ref{bosonic}) of the deformation obviously carries zero charge
under the R-current  (\ref{rcur}).  Instead, the fermionic piece (\ref{bilinear})
has zero charge only when the following conditions are satisfied
\begin{equation}
d_{\pm 1} = \pm a_j c_{\pm 1} \ .
\label{n2con}
\end{equation}
For a normalizable operator, which means we select $a_j=j+1$,
these conditions are satisfied automatically. As we will see soon, these
operators preserve also  supersymmetry.

\no
Of course it is expected that chiral or antichiral operators yield deformations
that preserve ${\cal N}=2$ supersymmetry. Indeed, a chiral primary state $|\chi\rangle$
satisfies $G^+_{-\frac{1}{2}}|\chi\rangle =0$ and hence the deformation
it yields is $|\widetilde \chi\rangle = G_{-\frac{1}{2}} |\chi\rangle = \frac{1}{\sqrt{2}}  G^-_{-\frac{1}{2}} |\chi\rangle$. This state is obviously
annihilated by $G^-_{-\frac{1}{2}}$ and furthermore, using the ${\cal N}=2$ commutation
relation $ \{G^+_{-\frac{1}{2}}, G^-_{-\frac{1}{2}}\}  = 2 L_{-1}$, we see that
it is also annihilated by  $G^+_{-\frac{1}{2}}$ up to a total derivative that does not
affect the action. Similarly, the sum of a chiral and an antichiral operator yields
again a ${\cal N}=2$ supersymmetric deformation. For instance the operator
with $j=m=0$
belongs to this category.
However, not all deformations preserving ${\cal N}=2$
supersymmetry need originate from a chiral or antichiral operator. For instance,
all operators with $|m|\neq (j+1)$ yield ${\cal N}=2$ preserving deformations
but none of them is chiral or antichiral primary.

\no
For non-normalizable operators with $a_j=-j$ there is only one
solution of the ${\cal N}=2$ constraints (\ref{n2con}) given by
  $j=m=0$. The corresponding operator is $\psi^3$ and it leads
  to the deformation $J^3+\psi^+ \psi^-$.  We hasten to point out that although this deformation
  preserves extended worldsheet supersymmetry, it does not lead
  to a spacetime supersymmetric background\footnote{
  Interesting applications of that mechanism of supersymmetry breaking
  in non-critical superstrings can be found in
   \cite{Suyama:2002xk, Itzhaki:2005zr, Harmark:2006sf}.}
   since it does not commute with the spacetime supercharges (\ref{spacesusy}).

 \no
Let us now check that the above argument, based on  R-charge neutrality, does not miss any solutions.
This can be done
by examining explicitly some terms of the OPE of $G^3(z)$ with the deforming
operator. Explicitly, this supercurrent is
\begin{equation}
G^3 = i \psi_3 \partial \phi - q \psi_\phi J_3 -q \psi_\phi \psi^+ \psi^-+
i q \partial\psi_3 + i \frac{q}{\sqrt{2}} (J^+ \psi^- - J^- \psi^+)\
\end{equation}
and let us keep only the terms of its OPE with the sum of
 (\ref{bosonic}) and (\ref{bilinear})
containing $\psi_3$. These terms read
\begin{equation}
\frac{iq}{z-w} \left (
a_j \sum_{r=-1}^1c_r \lambda_r \psi_3 J^r \Phi_{j;m-r}
 e^{-a_j q \phi}- a_j c_0  \psi_3 J^3 \Phi_{j;m}-
\frac{d_{+1}}{\sqrt{2}} \psi_3 J^+ \Phi_{j;m-1}
+\frac{d_{-1}}{\sqrt{2}} \psi_3 J^- \Phi_{j;m+1}
\right)
\end{equation}
and they vanish if and only if (\ref{n2con}) are satisfied. Hence, for the
class of operators under consideration,
the condition of vanishing R-charge is not only sufficient but also necessary
for preserving ${\cal N}=2$ supersymmetry.

 \subsubsection{${\cal N}=4$ supersymmetry}

\no
Similarly to the ${\cal N}=2$ case, a sufficient condition for preserving
${\cal N}=4$ SCFT invariance is that the deformation is a singlet under the 
corresponding R-symmetry
group $SU(2)_R$. The latter is generated by $J_R$ and two more generators
$S^\pm$:
\begin{equation}
SU(2)_R:\qq J_R=\psi^+\psi^-+\psi \psi^*, \quad S^+=\psi \psi^+, \quad S^-=\psi^-\psi^*\ .
\end{equation}
The OPEs of $S^\pm(z)$ with
(\ref{bilinear}) are zero provided that besides (\ref{n2con}),
which means that we already assume preservation of ${\cal N}=2$,
the following
condition is satisfied
\begin{equation}
d_0 = a_j c_0\ .
\label{n4con}
 \end{equation}

\no
This condition holds automatically  for all operators in the normalizable branch that
preserve ${\cal N}=2$. Let us present their fermionic pieces for completeness:
\begin{equation}
q N_j (j+1) \Big( i \sqrt{2} ( c_1 \psi^* \psi^+ \Phi_{j;m-1}+ c_{-1} \psi \psi^-
\Phi_{j;m+1}) + c_0 (-\psi \psi^* + \psi^+ \psi^-) \Phi_{j;m} \Big) e^{-q (j+1)\phi}\ .
\end{equation}
It is easy also to establish that these deformations preserve spacetime supersymmetry
by writing the above fermion bilinears  in bosonized form
 \begin{equation}
   \psi^* \psi^+ = e^{-i H_2 + i H_1}, \quad  \psi \psi^- = e^{i H_2 - i H_1},
   \quad
   -\psi \psi^* + \psi^+ \psi^- = -i \partial H_2 + i \partial H_1, \
 \end{equation}
where it is manifest that they commute with the spacetime supercharges
(\ref{spacesusy}). Actually since the same combination of $H_1$ and $H_2$
appears in $S^\pm$ and in the spacetime supercharges, we conclude that
any deformation preserving ${\cal N}=4$ superconformal
invariance automatically preserves spacetime supersymmetry as well.
As an example, we notice that the usual marginal deformation $J^3 \bar J^3$ 
of the bosonic $SU(2)$ WZW model can be promoted to an operator in the CHS background
that preserves ${\cal N}=(4,4)$ superconformal invariance in the following way
\begin{equation}
(J^3 - \psi \psi^* + \psi^+ \psi^-) (\bar J^3 - \bar \psi \bar  \psi^* + \bar \psi^+ \bar \psi^-) 
e^{-q \phi}\ .
\end{equation}

\no
Going now over to the non-normalizable sector, we observe that
the unique such operator
preserving ${\cal N}=2$, the one with $j=m=0$,  does not satisfy (\ref{n4con}) and hence
does not preserve ${\cal N}=4$. Hence, the non-normalizable deformation
$(J^3 + \psi^+\psi^-)(\bar J^3 + \bar \psi^+ \bar\psi^-) $ preserves
only  ${\cal N}=(2,2)$ supersymmetry (but not any spacetime supersymmetry
as we emphasized earlier).

\subsection{More supersymmetric deformations}

In this subsection we investigate the possibility that other classes of operators
lead to supersymmetric marginal deformations. One question is
under what conditions a deformation
that originates from the operator
\begin{equation} \label{gennseed}
\mu_3 \psi^3 \Phi_{j;m_3} e^{-a_j q \phi} +
\mu_+ \psi^+ \Phi_{j;m_+} e^{-a_j q \phi}+
\mu_- \psi^- \Phi_{j;m_-} e^{-a_j q \phi}\ ,
\end{equation}
preserves ${\cal N}=2$ and ${\cal N}=4$ superconformal invariance.
This operator differs from  $(\psi\Phi_j)_{j+1;m}
e^{-q a_j \phi}$
since the coefficients
$\mu_\pm, \mu_3$ are arbitrary and we do not assume a priori
any relation between $m_3$ and $m_\pm$.
As before, the deformation arises
by the action of $G_{-\frac{1}{2}} \bar G_{-\frac{1}{2}}$ on
 (\ref{gennseed}) and hence ${\cal N}=(1,1)$ supersymmetry is  guaranteed
 by construction. Notice that we consider a single $j$ since there cannot
 be any mixing among different $j$'s upon the action of the supercharges.

\no
The deformations that arise from each of the three operators
in (\ref{gennseed}) are:
\ba
&& q \mu_3 \Big(J_3 \Phi_{j;m_3} + i a_j \psi_\phi \psi^3 \Phi_{j;m_3} +
\psi^+ \psi^- \Phi_{j;m_3} +\frac{1}{\sqrt{2}} (j+m_3) \psi^+ \psi^3 \Phi_{j;m_3-1}
\nonumber\\
&&  \phantom{xxx} +\frac{1}{\sqrt{2}} (j-m_3) \psi^- \psi^3 \Phi_{j;m_3+1}\Big)e^{-a_j q \phi}\ ,\label{def1}
\ea
\begin{equation}
q \mu_+ \Big(\frac{1}{\sqrt{2}} J^+ \Phi_{j;m_+} + i a_j \psi_\phi \psi^+ \Phi_{j;m_+}
+ (m_++1)   \psi^3 \psi^+ \Phi_{j;m_+} +\frac{1}{\sqrt{2}} (j-m_+) \psi^-\psi^+
\Psi_{j;m_++1}\Big)e^{-a_j q \phi}\ ,\label{def2}
\end{equation}
and
\begin{equation}
q \mu_- \Big(\frac{1}{\sqrt{2}} J^-\Phi_{j;m_-} + i a_j \psi_\phi \psi^- \Phi_{j;m_-}
+(m_--1) \psi^3 \psi^- \Phi_{j;m_-} +\frac{1}{\sqrt{2}} (j+m_-) \psi^+\psi^-
\Phi_{j;m_--1}\Big)e^{-a_j q \phi}\ .\label{def3}
\end{equation}

\no
As explained previously, a sufficient condition for preserving ${\cal N}=2$ supersymmetry
is that these deformations are neutral under the $U(1)$ R-current $J_R$.
For the example under study
we find that neutrality under $J_R$ is guaranteed if the following conditions
are satisfied:
\begin{eqnarray}
\frac{1}{\sqrt{2}} \mu_3 (j+m_3) \Phi_{j;m_3-1} + \mu_+ (a_j-m_+-1)
\Phi_{j;m_+} &=& 0\,\label{n=21} \ ,
\nonumber\\
\frac{1}{\sqrt{2}} \mu_3 (j-m_3) \Phi_{j;m_3+1} -\mu_- (a_j+m_--1)
\Phi_{j;m_-} &=&0\ .
\label{n=22}
\end{eqnarray}
Furthermore,  in order to ensure ${\cal N}=4$ invariance we need to check that the deformations
are also neutral under the extra generators $S^\pm$ which, along with
$J_R$, generate the R-symmetry group $SU(2)_R$ of the ${\cal N}=4$
superconformal algebra. We find three conditions. Two of  those are identical with these
that guarantee ${\cal N}=2$ invariance. This is expected
since the $S^\pm$ generators close on $J_R$. The third  condition reads
\begin{equation}
\mu_3 (a_j-1) \Phi_{j;m_3} +\frac{1}{\sqrt{2}} \mu_+ (j-m_+) \Phi_{j;m_++1}
-\frac{1}{\sqrt{2}} \mu_- (j+m_-) \Phi_{j;m_--1} =0\ .
\label{n=4}
\end{equation}

 \no
  Equations (\ref{n=22}) and (\ref{n=4}) provide a set of sufficient
 conditions for the deformation (\ref{def1})+(\ref{def2})+(\ref{def3}) to preserve
 ${\cal N}=4$ supersymmetry. These conditions yield different equations
 for the coefficients $\mu_3 ,\mu_\pm$ depending on whether the charges
 $m_3$ and $m_\pm$ are related or not. The simplest case to analyze is
 that of $m_3=m_++1=m_- -1=m$. Then, the ${\cal N}=2$ conditions fix
 $\mu_\pm$ in terms of $\mu_3$ as
\be
\m_\pm = \mp {1\ov \sqrt{2}} {j\pm m\ov a_j\mp m }\ \m \ ,\qq \m_3=\m  \ .
\label{fhhf11}
\ee
The ${\cal N}=4$ condition yields a further constraint
\begin{equation}
\mu_3 (a_j-1) + \frac{1}{\sqrt{2}} \mu_+ (j-m+1) -\frac{1}{\sqrt{2}}
\mu_- (j+m+1) =0\ .
\label{n=4con}
\end{equation}
Upon combining with (\ref{fhhf11}) we find that in order to have non-trivial
solutions $a_j$ has to equal $a_j=-j,j+1,0$. In other words, the values of
$a_j$ that are singled-out by ${\cal N}=4$ supersymmetry include those
for which the deformation is marginal. Instead, the case of $a_j=0$ (with the
exception of $j=0$ which is analyzed below) leads to an irrelevant
operator.

 \no
For normalizable deformations, i.e.~$a_j=j+1$, the solution (\ref{fhhf11}) yields the
class of holographic operators
$(\psi \Phi_j)_{j+1;m} e^{-q(j+1)\phi}$. As we already know these operators
preserve ${\cal N}=4$ supersymmetry and hence it is not necessary to check
(\ref{n=4}) (it is automatically satisfied).
The purely bosonic part of the corresponding deformation is
\be
q\left(\m_3 J^3 \Phi_{j;m} + {\m_+\ov \sqrt{2}} J^+ \Phi_{j;m-1}
+ {\m_-\ov \sqrt{2}} J^- \Phi_{j,m+1}\right) e^{-q(j+1)\phi} \ ,
\label{fjkb1}
\ee
with $|m|\leqslant j+1$ and with the coefficients being given by
\be
\m_\pm = \mp {1\ov \sqrt{2}} {j\pm m\ov j+1\mp m }\ \m \ ,\qq \m_3=\m  \ .
\label{fhhf1}
\ee
Choosing $\mu= (j+1-m)(j+m+1)$ we obtain indeed the Clebsch--Gordan coefficients
\eqn{hglq}.

\no
For non-normalizable operators, i.e.~$a_j=-j$, the solution  (\ref{fhhf11})
boils down to
\begin{equation}
\m_\pm = \pm {1\ov \sqrt{2}}  \m \ ,\qq \m_3=\m  \ ,
\ee
which satisfies also the ${\cal N}=4$ condition (\ref{n=4con}).
This non-normalizable solution is by far more general than the one found in the previous
subsection, where $\mu_3$ and $\mu_\pm$ were specified
in terms of the Clebsch--Gordan  coefficients fixing $j=m=0$.
Instead, the current solution exists for any values of $j$ and $m$ that are allowed.
We will denote from now on the corresponding operator in (\ref{gennseed})
by $(\psi \Phi_j)_m e^{ qj \phi}$ since it does not have definite spin
(it does have definite ${\cal J}^3$ charge though).
The associated bosonic deformation reads
\be
q \mu \left( J^3 \Phi_{j;m} + {1 \ov 2} J^+ \Phi_{j;m-1}
- {1 \ov 2} J^- \Phi_{j;m+1}\right) e^{q j \phi}\ 
\label{fjk22}
\ee
and the fermion bilinear piece is
\begin{equation}
q j \mu \Big( i \psi \psi^- \Phi_{j;m+1} + i \psi^* \psi^+ \Phi_{j;m-1}+
(\psi \psi^* - \psi^+ \psi^-) \Phi_{j;m}\Big) e^{q j \phi}\ .
\label{nonholofer1}
\end{equation}
Since these deformations preserve ${\cal N}=4$ superconformal invariance,
they also preserve  
spacetime supersymmetry. Notice that
these operators, for generic $j$
and $m$, do not have a holographic counterpart since they do not have
definite total spin,  and therefore  cannot
correspond to an LST deformation. Their interpretation
in terms of NS5-branes will be uncovered in section 4.

\no
There are two more classes of operators that lead automatically to
${\cal N}=2$ preserving deformations. These are  $\psi \Phi_{j;-j} e^{-q (j+1)} \phi$ and
 $\psi^* \Phi_{j;j} e^{-q (j+1)}
$ which, as was shown in subsection 2.2, are chiral and antichiral
primaries, respectively. They were not captured by the  analysis we just performed
since the ansatz (\ref{gennseed}) does not contain the fermion $\psi_\phi$.
 It can be checked that
the corresponding deformations preserve also ${\cal N}=4$ supersymmetry
and therefore spacetime supersymmetry.
The purely bosonic piece of the deformation coming from  $\psi \Phi_{j;-j} e^{-q (j+1)} \phi$
is
\begin{equation}
\frac{i}{\sqrt{2}} (\partial\phi + q J_3) \Phi_{j;-j} e^{-q(j+1)\phi}
\end{equation}
and the fermion bilinear piece is
\begin{equation}
\frac{q}{\sqrt{2}} \left(i \psi^+ \psi^-  \Phi_{j;-j} - i \psi \psi^* \Phi_{j;-j} +
2 j \psi^- \psi \Phi_{j;-j+1}\right)  e^{-q(j+1)\phi}\ .
\end{equation}
 
 \no
Notice that  $\psi \Phi_{j;-j} e^{-q (j+1)} \phi$ and  $\psi^* \Phi_{j;j} e^{-q (j+1)}$
do not have definite spin under the spacetime symmetry
$SO(4)$ and do not appear independently in the holographic dictionary
 (only their imaginary part for $j=0$, which is
 $\psi_3 e^{-q \phi}$, does have a holographic interpretation).
 In that respect, they are similar to the non-normalizable operators
 we discussed earlier, which also preserve worldsheet and spacetime
 supersymmetry. Although both classes of operators leave intact the 6-dim Lorentz invariance
associated with the worldvolume of the NS5-branes,
they lack an interpretation in LST.


\no
 The non-normalizable versions of
$\psi \Phi_{j;-j} e^{-q (j+1)} \phi$ and  $\psi^* \Phi_{j;j} e^{-q (j+1)}$
are chiral
and antichiral, respectively, but not primary. They are examples of operators where
$h=|q|/2$ but due to non-unitarity they fail to be chiral primary. Furthermore,
it can be checked that they preserve ${\cal N}=2$ supersymmetry
but not ${\cal N}=4$ and hence they break spacetime supersymmetry.

\no
 An interesting observation is that the extra operators
$\psi \Phi_{j;-j} e^{-q (j+1)} \phi$ and  $\psi^* \Phi_{j;j} e^{-q (j+1)}$
are actually BRST trivial
in the ${\cal N}=2$  topologically twisted theory. The reason is that they
arise from the action of $G^+(z)$ on $\Phi_{j;-j} e^{-q(j+1)\phi}$ since
\begin{equation}
G^+(z) \Phi_{j;-j} e^{-q(j+1)\phi} (w) \sim \frac{i q(2j+1)}{z-w} \psi \Phi_{j;-j} e^{-q(j+1)\phi} (w)
\end{equation}
(and similarly for the complex conjugate). In the topological theory where
the energy momentum tensor is $T(z)+\frac{1}{2} \partial J_R(z)$
the BRST charge is $Q_{BRST} = \oint G^+(z) dz$ and
$\psi \Phi_{j;-j} e^{-q(j+1)\phi}$ is a trivial element of the BRST cohomology.

\no
Let us also point out that the holographic operators $(\psi \Phi_j)_{j+1;m} e^{-q(j+1)\phi}$
originate from the action of the extra two ${\cal N}=4$  supercharges on
 $\Phi_{j;-j} e^{-q(j+1)\phi}$. These supercurrents read
\begin{equation}
\begin{array}{rcl}
\displaystyle{\tilde G^+}&=&\displaystyle{i
\psi^+\big(\partial\phi+q J_3- q \psi\psi^*\big)+i q
\partial\psi^+-
 q J^+\psi} \ , \\
\displaystyle{\tilde G^-}&=&\displaystyle{i
\psi^-\big(\partial\phi-q J_3 + q \psi\psi^*\big)+i q
\partial\psi^--
 q J^-\psi^* } \ ,
\end{array}
\end{equation}
and, along with $G^\pm(z)$, generate the ${\cal N}=4$ superconformal algebra. Then
it holds that
\begin{equation}
\tilde G^+(z) \Phi_{j;j} e^{-q(j+1)\phi} (w) \sim \frac{i q(2j+1)}{z-w} \psi^+ \Phi_{j;j}
e^{-q(j+1)\phi} (w)\ .
\end{equation}
Hence, with respect to the ${\cal N}=2$ algebra generated
by $\tilde G^{\pm}(z)$, the operator $\psi^+ \Phi_{j;j} e^{-q(j+1)\phi}$
would be BRST trivial after the topological twisting. However, neither
$\psi \Phi_{j;-j} e^{-q(j+1)\phi}$ nor $\psi^+ \Phi_{j;j} e^{-q(j+1)\phi}$  are trivial
as elements of the BRST cohomology of the ${\cal N}=4$ topological string.

\no
 When  $j=m=0$
the real part of $\psi \Phi_{j;-j} e^{-q(j+1)  \phi}$ and of its non-normalizable version
preserves ${\cal N}=4$ but the deformation it leads to, whose purely bosonic piece
reads $\partial\phi\bar \partial\phi e^{-q a_0 \phi}$,
is trivial since
it is tantamount to a coordinate redefinition of the linear dilaton direction.
This triviality, however, does not seem to persist when $j\neq0$
since primaries of the $SU(2)_{k-2}$ WZW model couple to the derivatives
of the dilaton.

\subsection{Comments and summary}

Normalizable CFT operators of the form $(\psi\bar \psi \Phi_j)_{j+1;m, \bar m} e^{-q(j+1)\phi}$
correspond holographically to VEVs of the operators
$ \widetilde {\rm tr} (X^{i_1}X^{i_2}\cdots X^{i_{2j+2}})$ that parametrize
the moduli space of LST. Notice that spacetime supersymmetry does not change as
we move  in the moduli space since any configuration of parallel
NS5-branes, irrespectively of  their positions in the transverse space,
preserves 16 supercharges in type II theories. Consequently,  the ${\cal N}=4$ superconformal
symmetry
of the original underlying CFT should also be left intact \cite{Banks:1988yz}.
We have shown that
all deformations
originating from $(\psi \bar \psi \Phi_j)_{j+1;m, \bar m} e^{-q(j+1)\phi}$
 preserve both
${\cal N}=(4,4)$ worldsheet supersymmetry and 16 spacetime supercharges.
Therefore, all of those that
bear non-vanishing couplings
$\lambda_{j;m,\bar m}$ can be in principle present in the deformed
Lagrangian.

\no
The last observation is particularly puzzling for two reasons.
First, as noticed in  \cite{Aharony:2003vk},
there is a mismatch between the
number of couplings $\lambda_{j;m,\bar m}$, which grows as $k^3$, and the number
of parameters that determine a point in the moduli space of LST, the latter being
$4(k-1)$. Second, \cite{Fotopoulos:2007rm} established that
the most general
planar deformation of the NS5-branes\footnote{Notice that
\cite{Fotopoulos:2007rm} considered deformations of a circular distribution
of NS5-branes where the underlying CFT is $SL(2,\mathbb{R})_k/U(1) \times SU(2)_k/U(1)$.
As we mentioned already, 
 there is a 1-1 correspondence between operators in that theory and the CHS theory
 studied here, so that 
all results pertaining to $SL(2,\mathbb{R})_k/U(1) \times SU(2)_k/U(1)$ deformations
can be rephrased in the CHS theory.}
 was captured by a subset of the possible
deforming operators, namely those that are (chiral, chiral) primaries as well
as their (antichiral, antichiral) conjugates. It is also quite straightforward to see
that non-planar deformations of the NS5-brane are captured by (chiral, antichiral)
and (antichiral, chiral) operators.

\no
Therefore, it seems that the chiral ring operators,
namely those with $m, \bar m =\pm(j+1)$ , are sufficient to capture the most
general geometric NS5-brane deformation. Then, the puzzle raised in
  \cite{Aharony:2003vk} is resolved since the number of
such operators is precisely $4(k-1)$ (recall that $j=0,\frac{1}{2},\ldots,\frac{k-2}{2}$
and we have to combine the holomorphic with the antiholomorphic part).
This is also in line with the fact that in the T-dual theory, which is
described by a $\sigma$ model with an ALE target space and without
the complications due to the presence of NS-NS flux,
only operators in the chiral ring correspond to geometric moduli.
Therefore, we will also dub ``non-holographic`` all operators of the form
 $(\psi \bar \psi \Phi_j)_{j+1;m, \bar m} e^{-q(j+1)\phi}$ with
 $|m|$ or $|\bar m|$ different than $j+1$.

\no
One extra argument in support of this proposal is
that the marginal deformations originating from chiral (or antichiral) operators are actually
exactly marginal. As we said earlier, the reason is that these operators have
protected conformal dimensions since the latter are fixed in terms
of the non-renormalized $U(1)$ R-charge as $h=\frac{q}{2}$. Since the NS5-branes
can be finitely separated without spoiling the
conformal invariance of the worldsheet theory, we are lead to the conclusion
that only exactly marginal deformations, which in principle can be integrated
to finite deformations, should be used to perturb the original CFT.

\no
Notice that deformations originating from non-chiral operators are not, in general,
exactly marginal since they do not satisfy the criterion of \cite{Chaudhuri:1988qb}.
However, there is an exception provided by the operator $j=m=\bar m=0$. The purely posonic
piece of the corresponding deformation is
 $J^3 \bar J^3 e^{-q \phi}$ and it is well-known  \cite{Chaudhuri:1988qb}
 that $J^3 \bar J^3$ is an
exactly marginal operator of the $SU(2)_{k-2}$ WZW model. The Liouville dressing does not
modify the argument of \cite{Chaudhuri:1988qb} since the operator
$e^{-q \phi}$ is equivalent to the identity operator in Liouville theory
and its OPE with itself is trivial.

\no
Therefore, we propose that the normalizable operator with $j=m=\bar m =0$
should also be taken into account
when  one considers NS5-brane deformations. An additional reason for doing so
is that in the simple example where the point-like configuration of NS5-brane
is deformed to a circle, this operator yields the leading deformation of
the CHS theory (or, more precisely, of  its T-dual) towards the
model  $SL(2,\mathbb{R})_k/U(1) \times SU(2)_k/U(1)$  \cite{Aharony:2003vk}
(see also subsection 3.3 for more details).

\no
Finally, let us point out that the fact that there are no
non-normalizable operators in the holographic dictionary that
preserve the ${\cal N}=4$ superconformal
symmetry, and consequently the full worldvolume supersymmetry,
ties nicely with the fact that there should not exist any such deformations of the 
5+1-dimensional LST.  The other class of
non-normalizable operators
\be
q \mu \left( J^3 \Phi_{j;m} + {1 \ov 2} J^+ \Phi_{j;m-1}
- {1 \ov 2} J^- \Phi_{j,m+1}\right) e^{q j \phi},
\ee
that preserves ${\cal N}=4$, does not correspond to an LST deformation but,
as we will see in more detail in section 4, moves us
 away from the NS5-brane horizon.

\no
The salient findings of this section
are summarized for convenience in the table below.
Notice that we have not included the complex conjugates of the chiral primaries
which are antichiral and they preserve the same amount of supersymmetry
as their chiral counterparts. Operators with unspecified $j$ and $m$ labels
are assumed to be generic, i.e.~not for the cases $j=m=0$ and/or $m=\pm(j+1)$
when they reduce to other operators present in the table.

\begin{center}
  \begin{tabular}{   | c|| c | c  | c | c |  c |}
    \hline
  {\bf operator}    & {\bf chiral}  & {\bf primary} &   {\bf ${\cal N}=2$ }&  {\bf ${\cal N}=4$ }
  & {\bf spacetime susy} \\ \hline\hline\hline
    $\psi^+ \Phi_{j;j} e^{-q (j+1)\phi}$ & $\surd$ & $\surd$   &$\surd$ & $\surd$ & $\surd$\\ \hline
     $\psi^+ \Phi_{j;j} e^{q j \phi}$ &   & $\surd$  &  & &  \\ \hline\hline
     $ (\psi \Phi_j)_{j+1;m} e^{-q (j+1)\phi}$ & & &  $\surd$   &$\surd$  & $\surd$  \\ \hline
      $ (\psi \Phi_j)_{j+1;m} e^{q j \phi}$ & & &     &  & \\ \hline\hline
      $\psi_3 e^{-q \phi}$ &   &  $\surd$   & $\surd$ & $\surd$   & $\surd$ \\ \hline
        $\psi_3 $ &  &   & $\surd$ &  &  \\ \hline\hline\hline
        $(\psi \Phi_{j})_m e^{-q (j+1) \phi}$ &   &     &  &  & \\ \hline
       $(\psi \Phi_{j})_m  e^{q j \phi}$ &     &  &  $\surd$  & $\surd$  & $\surd$ \\ \hline\hline
     $\psi \Phi_{j;-j} e^{-q (j+1) \phi}$ & $\surd$  &  $\surd$   & $\surd$ & $\surd$   & $\surd$ \\ \hline
       $\psi \Phi_{j;-j} e^{q j \phi}$ & $\surd$    &  &  $\surd$  &  &  \\ \hline\hline
    $\psi_\phi e^{-q \phi}$ &   &  $\surd$   & $\surd$ & $\surd$  & $\surd$ \\ \hline
        $\psi_\phi $ &  &   & $\surd$ & $\surd$ & $\surd$ \\ \hline\hline  
     \end{tabular}
\end{center}

\section{\boldmath Deformations of $SU(2)_k \times \mathbb{R}_\phi$ and NS5-branes
  \unboldmath}

In this section we will analyze several configurations of NS5-branes
using the holographic correspondence  (\ref{holodic}) and its refinement
proposed in the previous section.
The configurations under study
will be thought of as small deformations of a stack of NS5-branes put
at the point $x^6=x^7=x^8=x^9=0$. Accordingly,  the exact CFT
$SU(2)_k \times \mathbb{R}_\phi$ describing the latter is deformed and
we will show how several physical features of the configurations of NS5-branes
under study can be inferred from the analysis of the corresponding CFT deformations.

\subsection{Generalities}

\no
We revisit now the holographic dictionary (\ref{holodic}) and explain how it works
in detail. Since there
are no $m$ and $\bar m$ indices at the left side, an obvious question is
how these charges are determined in terms of the indices $i_1,\ldots,i_{2j+2}$
for a given LST operator.
As shown in \cite{Aharony:2003vk}, this can be done by using a parametrization
of the moduli space in terms of two complex variables that span
the two orthogonal hyperplanes transverse to the NS5-branes:
\begin{equation}\label{defAB}
A\equiv X^6+i X^7, \qq
B\equiv X^8+i X^9\ .
\end{equation}
Embedding the rotational $SO(2)_A \times SO(2)_B$ of the $A$ and $B$ planes
in the $SU(2)_L \times SU(2)_R$ symmetry of the CHS background  so
that $SO(2)_A$ is generated by ${\cal J}_3-\bar {\cal J}_3$ and
$SO(2)_B$ is generated by ${\cal J}_3+\bar {\cal J}_3$, leads to the following
charge assignments
\begin{equation}
m_A=\frac{1}{2},\qq \bar m_A=-\frac{1}{2},\qq m_B=\frac{1}{2},\qq \bar m_B=\frac{1}{2}
\ .
\end{equation}

\no
Subsequently, the general recipe (\ref{holodic}) takes a more precise form 
\begin{equation}\label{holodic2}
\widetilde {\rm  tr} \Big(A^x B^y (A^*)^z (B^*)^w
\Big) \longleftrightarrow (\psi\bar \psi\Phi_j)_{j+1;m,\bar m} e^{-q(j+1)\phi}\ ,
\end{equation}
where $-(j+1)\leqslant m,\bar m \leqslant (j+1)$
and the positive powers $x,y,z,w$ are related to the charges $j,m,\bar m$ as
\begin{equation}
x+y=j+1+m, \quad
z+w =j+1-m, \quad
y+z=j+1+\bar m, \quad
w+x=j+1-\bar m\ .
\end{equation}
The corresponding couplings $\lambda_{j;m,\bar m}$ are given by
 \begin{equation}
 \lambda_{j;m,\bar m} = \frac{1}{k}
 \widetilde {\rm  tr} \Big(A^x B^y (A^*)^z (B^*)^w \Big)  \label{coupl}
 \end{equation}
 and symmetrization is not necessary since
we are interested in points in the LST moduli space where
$A$ and $B$ are diagonal. The $1/k$ factor is introduced so that the couplings
are ${\cal O}(1)$ in general. Furthermore, one should keep only the traceless
combinations in (\ref{coupl}).

 \no
  The analysis of the previous section indicated that generically we should consider
  only the
 operators that are either chiral or antichiral as well as the operator with $j=m=\bar m=0$.
The associated couplings are 
\begin{eqnarray}
\lambda_{j;j+1,j+1} &=& \frac{1}{k} \widetilde {\rm tr} (B^{2j+2})\ , \quad
\lambda_{j;-j-1,-j-1} =\frac{1}{k} \widetilde  {\rm tr} \big((B^*)^{2j+2}\big)\ , \\
\lambda_{j;j+1,-j-1} &=&\frac{1}{k} \widetilde {\rm tr} (A^{2j+2})\ , \quad
\lambda_{j;-j-1,j+1} = \frac{1}{k} \widetilde {\rm tr} \big((A^*)^{2j+2}\big)\ , 
\end{eqnarray}
which are automatically traceless, and 
 \begin{equation}
 \lambda_{0;0,0} = \frac{1}{k} \widetilde {\rm tr} ( B B^* - A A^*)\ ,
\label{coupj0m0}
 \end{equation}
 where the relative sign is chosen so that it is traceless.

 \subsection{NS5-branes on a 3-sphere}

The first configuration we would like to consider is that
of  a continuous distribution of NS5-branes on an $S^3$ of radius $R$
embedded in the transverse $\mathbb{R}^4$. This configuration is
described by
\begin{equation}
A= R \cos\theta e^{i \phi}, \quad
B=R \sin\theta e^{i \tau}\ ,
\label{conts3}
\end{equation}
where $\theta \in [0,\pi/2), \phi \in [0, 2\pi), \tau \in [0,2\pi).$ We would like to approximate
this distribution by a sequence of discrete ones containing $k$ NS5-branes
so that the limit $k\rightarrow \infty$ yields (\ref{conts3}).
Then, each of the coordinates $\theta,\phi,\tau$ on the sphere is
discretized in terms of an index $a,b,c$ as follows
\begin{eqnarray}
\sin^2\theta &=& \frac{a}{k_1}\ , \qq a=0,\ldots,k_1\ ,
\nonumber
\\
\phi&=& \frac{2 \pi b}{k_2}\ , \qq b=0,\ldots,k_2\ , \\
\tau&=& \frac{2 \pi c}{k_3},\qq c=0,\ldots,k_3
\nonumber
\ .
\end{eqnarray}
We can verify that
\begin{equation}
\frac{1}{2\pi^2} \int \sin\theta \cos\theta d\theta d\phi d\tau = \frac{1}{k}
\int da db dc \ ,
\end{equation}
so that the total number of NS5-branes is $k=k_1k_2 k_3$. Notice
that we assume that the discretization is smooth and hence that $k_1,k_2$
and $k_3$  are large.
The discrete distribution is described by  the $k \times k$ matrices
\begin{equation}
A_{a,b,c} =  R \sqrt{1-\frac{a}{k_1}} e^{\frac{2 \pi i b}{k_2}} \mathbb{I}_c\ , \qq
B_{a,b,c} = R \sqrt{\frac{a}{k_1}} \mathbb{I}_b e^{\frac{2 \pi i c}{k_3}}\ .
\end{equation}
By definition we have
 ${\rm tr} (A) \equiv \sum_{a=0}^{k_1} \sum_{b=0}^{k_2} \sum_{c=0}^{k_3}
A_{a,b,c}$ and we conclude that
${\rm tr} (A) ={\rm tr}( B)=0$ as it should.

\no
Before considering the chiral and antichiral operators
let us see what happens with the $j=m=\bar m=0$ operator. Its coupling
turns out to be zero since
 \begin{equation}
 \sum_{a,b,c}  \left(B_{a,b,c} B^*_{a,b,c} -A_{a,b,c} A^*_{a,b,c} \right) \sim
\sum_{a=0}^{k_1} (\frac{2}{k_1} a -1) =0\ .
\end{equation}
Hence, for this particular configuration this operator does not
appear in the perturbed theory. Notice that we have used the standard trace
since, as argued in \cite{Aharony:2004xn}, the multi-trace corrections are
subleading when $k$ is large and $j$ is finite. For the same reason
we will employ the usual single trace in the computation of $\lambda_{j;m,\bar m}$
below, since the values of $j$ that we will consider will be large, but finite.

\no
We proceed now with the computation of the coefficients $\lambda_{j;m,\bar m}$ for the cases
where $m=\pm (j+1)$ and $\bar m= \pm (j+1)$. We have
\begin{eqnarray}
\lambda_{j;m,\bar m} &=& R^{|m-\bar m|+|m+\bar m|}
\frac{1}{k} \sum_{a=0}^{k_1} \sum_{b=0}^{k_2} \sum_{c=0}^{k_3}
 \left(1-\frac{a}{k_1}\right)^{\frac{|m-\bar m|}{2}}
 \left(\frac{a}{k_1}\right)^{\frac{|m+\bar m|}{2}}
e^{\frac{2 \pi i b}{k_2}(m-\bar m)} e^{\frac{2 \pi i c}{k_3}(m+\bar m)}
\nonumber\\
&=& R^{|m-\bar m|+|m+\bar m|} \frac{k_2 k_3}{k} \sum_{a=0}^{k_1} \left(1-\frac{a}{k_1}\right)^{\frac{|m-\bar m|}{2}} \left(\frac{a}{k_1}\right)^{\frac{|m+\bar m|}{2}} \delta_{m-\bar m,0 \; {\rm mod} \; k_2}  \delta_{m+\bar m,0 \; {\rm mod}
\; k_3} \ . \nonumber
\end{eqnarray}
For large $k_1, k_2, k_3$ we can approximate the summation over $a$ with an integral.
We get
\begin{equation}
\lambda_{j;m,\bar m} = R^{|m-\bar m|+|m+\bar m|} B\left(1+\frac{|m-\bar m|}{2}, 1+\frac{|m+\bar m|}{2}\right)
\delta_{m-\bar m,0 \; {\rm mod} \; k_2}  \delta_{m+\bar m,0 \; {\rm mod}
\; k_3} \ ,
\end{equation}
where $B(x,y)$ is the Euler beta function. Notice that if we had made
the same approximation for the summations over $b$ and $c$ we would have found
that only the coefficient with  $m=\bar m=0$ is non-zero, therefore missing
all other possibilities.

\no
Let us now try to understand the behavior of the coefficients
$\lambda_{j;m,\bar m}$ for operators that are either
(chiral, chiral) or (chiral, antichiral). The other two cases are related to these
two by conjugation.  In the first case we have $(m,\bar m)=(j+1,j+1)$
and the corresponding coupling is
\begin{equation}
\lambda_{j;j+1,j+1} = \frac{R^{2j+2} }{j+1} \delta_{2j+2,0 \; {\rm mod}
\; k_3} \ .
\end{equation}
Since $2j+2$ goes up to $k$, there are $k/k_3=k_1 k_2$ values of
$j$ which give non-vanishing $\lambda_{j;j+1,j+1}$. For the
(chiral, antichiral) case, where $(m,\bar m)=(j+1,-j-1)$, the coupling is
\begin{equation}
\lambda_{j;j+1,-j-1} = \frac{R^{2j+2} }{j+1} \delta_{2j+2,0 \; {\rm mod}
\; k_2}
\end{equation}
and we have $k/k_2=k_1 k_3$ values of $j$ that yield non-vanishing coefficients.

\no
We can now consider the purely bosonic deformation corresponding to the (chiral, chiral)
operators:
\begin{equation}
\sum_{j=0}^{\frac{k-2}{2}} \frac{q^2}{2} \lambda_{j;j+1,j+1}  J^+ \bar J^+
\Phi_{j;j,j} e^{-q (j+1) \phi}\ .
\end{equation}
Explicitly this reads
\ba
&& \sum_{j=0}^{\frac{k-2}{2}} \frac{1}{k} \frac{R^{2j+2}}{j+1}
 J^+ \bar J^+
\Phi_{j;j,j} e^{-q (j+1) \phi}  \;   \delta_{2j+2,0 \; {\rm mod}
\; k_3}
\nonumber\\
&& \phantom{xxxxxxxx} =\
\sum_{j=0}^{\frac{k-2}{2}}  \frac{J^+ \bar J^+ }{k(j+1)}
 \Phi_{j;j,j}\ e^{-q (j+1) (\phi-\sqrt{2 k} \ln R )}
 \;   \delta_{2j+2,0 \; {\rm mod}
\; k_3} \ .
\ea

\no
We start the analysis by noticing that
the smallest value of $j$ that contributes is of order $k_3$ and hence it is large.
Consequently,
if $\phi - \sqrt{2k} \ln R = x < 0$ the exponential is
$e^{- \frac{j}{\sqrt{k}} \; x}$ and for $j$ of order $k_3$ it creates a potential wall
that does not allow penetration in the $x<0$ region.
In terms of the linear dilaton coordinate this wall is located at  $\phi_0 = \sqrt{2k} \ln R$.
The region $x>0 \Leftrightarrow \phi > \phi_0$ has a potential that goes very rapidly
to zero as  $k\rightarrow \infty$.
Hence, when $k \rightarrow \infty$ we obtain the original $SU(2)
\times \mathbb{R}_\phi$ theory,
but with a truncated dilaton direction $\phi > \phi_0$. In terms of the usual radial
coordinate the wall is located exactly at the radius of the 3-sphere.

\no
What we have just described is a stringy way of creating an impenetrable domain for
all modes in the theory. This should be compared to a purely gravitational approach in which
one postulates, without providing a microscopic origin, that the region $r<R$ is not part
of the space and  $\d$-function source terms are added so that 
the equations of motion are satisfied at $r=R$.

\no
This analysis is in perfect agreement
with the application of the Gauss law for a configuration
of NS5-branes spread on
 a 3-sphere. In the limit of large $k$ the $SO(4)$ symmetry is restored
and according to the Gauss law, outside of the 3-sphere we should obtain the same solution
as that of a point-like configuration of NS5-branes, i.e.~$SU(2)_k \times \mathbb{R}_\phi$,
but for the fact that the dilaton direction is truncated to $\phi > \phi_0$. Instead,
inside the sphere we have no
sources and the solution should be the free CFT corresponding to four
free bosons parametrizing $\mathbb{R}^4$.
However, this effect cannot be seen explicitly in our analysis
since the perturbation blows up
and cannot be considered as a small deformation of the original CFT.

 \subsection{NS5-branes on a circle: redux}

We consider now a configuration of $k$ NS5-branes arranged symmetrically
on a  polygon inscribed in a circle of radius R in the B plane.  This is a configuration
that has been discussed extensively in the literature since it admits an exact
CFT description. It is described by
the $k \times k$ matrices
 \begin{equation}
 A_a=0, \quad B_a = R e^{2 \pi i a/k}\ ,
 \end{equation}
so that the only non-zero couplings are $\lambda_{0;0,0} = R^2$
and $\lambda_{\frac{k-2}{2};\frac{k}{2},\frac{k}{2}}\lambda_{\frac{k-2}{2};-\frac{k}{2},-\frac{k}{2}} R^k$. Notice that ${\rm tr} (B^l) = 0$ when $l<k$ and
 hence there is no difference between the usual trace and the one
 containing multi-traces.

 \no
 The operator
corresponding to the second coupling behaves exactly as the operators
discussed in the previous subsection, i.e.~it rapidly vanishes when we are probing
the region outside the ring. Hence, the theory is modified only by the presence
of the operator corresponding to  $\lambda_{0;0,0} = R^2$, which is
$\psi_3 \bar\psi_3 e^{-q \phi}$ and whose  bosonic part reads
 $J_3 \bar J_3 e^{-q \phi}$.
This operator drives the deformation
\begin{equation}
SU(2)_k \times \mathbb{R}_\phi = \displaystyle \frac{SU(2)_k}{U(1)}  \times U(1) \times
\mathbb{R}_\phi  \longrightarrow
\displaystyle \frac{SU(2)_k}{U(1)}\times \frac{SL(2,\mathbb{R})_k}{U(1)}\ ,
\end{equation}
where a T-duality has been also been performed in the first step.
 The latter model is indeed well-known to provide the exact CFT description of a
 circular configuration of NS5-branes in the near-horizon limit
 \cite{Sfetsos:1998xd}.
 
 \no
We may verify explicitly that statement from the expressions for the corresponding
background \cite{Sfetsos:1998xd}
\ba
ds^2 & = & k\left[d\rho^2+ d\th^2+{1\ov \S}(\tanh^2 \r\
d\tau^2 + \tan^2\th\ d\psi^2)\right]\ ,
\nonumber\\
 B_{\tau\psi}& = &{k\ov \S}\ ,
\label{meci}\\
 e^{-2\Phi} & = & \S \cosh^2\r \cos^2\th\ ,
\nonumber
\ea
where the dilaton is included for completeness and
\be
\S=\tanh^2\r \tan^2\th+1\ .
\ee
Indeed,
expanding \eqn{meci} for large $\r$ which, effectively, is equivalent to a circle
of small size, we get that
\ba
ds^2 & =  & ds_{(0)}^2 + 4 e^{-2 \r}(\sin^4\th \ d\psi^2 - \cos^4 \th \ d\tau^2)
+ {\cal O}\left(e^{-4 \r}\right)\ ,
\nonumber\\
B_{\tau\psi} & = & B_{\tau\psi}^{(0)} + 4 e^{-2 \r} \sin^2\th \cos^2\th + { \cal O}\left(e^{-4 \r}\right)\ ,
\ea
where the fields indexed with a zero correspond to the $SU(2)$ WZW unperturbed case.
It is now straightforward to verify that the perturbation is just $-J_3\bar J_3 e^{-q\phi}$, where
$\phi=\sqrt{2 k}\ \r$ and
\be
J_3 =2 ( \cos^2\th \del \tau + \sin^2\th \del \psi)\ ,
\qq
\bar J_3 =2 ( \cos^2\th \bar \del \tau - \sin^2\th \bar \del \psi)\ ,
\ee
are the chirally and antichirally conserved Cartan currents.

\subsection{NS5-branes on orthogonal circles}

Another interesting configuration is that of $k$ NS5-branes put on two circles
of the same radius R on the two planes A and B. This is described by the
$k' \times k'$ matrices
\begin{equation}
 A_a=R e^{2 \pi i a/k'}, \quad B_a = R e^{2 \pi i a/k'}\ ,
 \end{equation}
with $k'=k/2$. In this case the only non-zero couplings are
$\lambda_{\frac{k'-2}{2};\pm \frac{k'}{2},\pm \frac{k'}{2}}=\frac{1}{2} R^{k'}$.
As in the previous case of NS5-branes put on a single circle, there is no difference
between the usual trace and the tilde one.
For large $k'=k/2$ the corresponding operators vanish rapidly in the region
outside of the two rings and since there are no other non-vanishing operators
(unlike the case of one ring where $\lambda_{0;0,0} \neq 0$) the situation
resembles that of the 3-sphere.

\no
The previous remark leads to a puzzle, since the solutions corresponding to
the 3-sphere and the two circles are obviously different even in the
$k \rightarrow \infty $ limit. One can understand why the two configurations
behave similarly by examining the multipole expansion of the corresponding
harmonic functions, in conjunction also with a similar expansion for the case
of one circle.
The reason is that on physical grounds the first non-vanishing multipole moment
triggers the leading perturbation of the original CFT.
Since the 3-sphere configuration behave as a point-like
charge, all of its multipole moments vanish by definition. The dipole moments
$p^i = \int d^4 x \rho(x) x^i $, where $i=6,7,8,9$ is a vector index in
the 4-dimensional transverse $\mathbb{R}^4$ and $\rho(x^i)$ is
the density of NS5-branes, are zero for both the case of one and two circles.
This is easy to check using the normalized
 densities
 \begin{equation}
 \rho_{\rm 1-circ.}(x)=\frac{1}{\pi} \delta\left(R^2-(x^8)^2-
(x^9)^2\right) \delta(x^6) \delta(x^7)
\end{equation}
and
\begin{equation}
\rho_{\rm 2-circ.}(x)=\frac{1}{2\pi} \Big[\delta\left(R^2-(x^8)^2-
(x^9)^2\right) \delta(x^6) \delta(x^7) + \delta\left(R^2-(x^6)^2-
(x^7)^2\right) \delta(x^8) \delta(x^9) \Big]\ .
\end{equation}
 Now, one can further check that the quadrupole moments
\be
Q^{ij} =  \int d^4 x \rho(x) \left(x^i x^j -\frac{1}{4} \delta^{ij} x^2\right)\ ,
\ee
vanish for the two circles, rendering the solution similar to that of the 3-sphere up to this order.
Instead, the single circle behaves differently since
 it has non-vanishing quadrupole moments, for instance
 $Q_{66}=Q_{77}=-\frac{1}{4}
R^2$.  More generally, the quadrupole moments vanish for every distribution
that is identical on the A and B planes and which has no dependance on the
angular coordinate of the plane.

\subsection{NS5-branes on a line and symmetry considerations}

A final configuration we would like to consider is that of NS5-branes put on
a line, for instance in the B plane and along the $x^8$ direction. In that case
we have $A=0$ and $B=B^*$ and all couplings $\lambda_{j;j+1,j+1}$, their conjugates
as well as $\lambda_{0;0,0}$ are generically non-zero. Notice that our discussion
here is independent of the actual distribution on the line.
The configuration we consider
is invariant under an $SO(3)$ group of transverse symmetries and a natural question
is how this symmetry manifests itself in the CFT deformations.

\no
Before tackling this
problem, let us start with a generic configuration of NS5-branes in the transverse
$\mathbb{R}^4$ where both $A$ and $B$ are non-zero.
The various cases will be discussed in reference to fig.~1 below which summarizes and
depicts them geometrically.
Then, generically, all the couplings
$\lambda_{j;\pm (j+1),\pm(j+1)}$ and $\lambda_{0;0,0}$ are non-zero
and the $SO(4) = SU(2)_L \times SU(2)_R $ symmetry is completely broken.
Instead, an arbitrary deformation on a single plane should preserve
the $SO(2)$ symmetry associated with rotations in the plane orthogonal to the first one.
For instance, spreading the branes in the $B$ plane triggers the
(chiral, chiral) and (antichiral, antichiral) operators corresponding to the
couplings $\lambda_{j;j+1,j+1}$ and $\lambda_{j;-j-1,-j-1}$ respectively, as
well as $\lambda_{0;0,0}$. The purely bosonic pieces of the associated deformations
are
\be
J^+ \bar J^+ \Phi_{j;j,j} e^{-q(j+1)\phi}\ , \qq J^- \bar J^- \Phi_{j;-j,-j}e^{-q(j+1)\phi}\ ,
\qq J^3 \bar J^3e^{-q(j+1)\phi}\ .
\ee
All these operators are invariant\footnote{
As usual, when we refer to the action of the currents
we mean the action of their zero-modes and hence what we should check is the
simple pole in their OPE with the operator under consideration.}
under the generator
$J^3-\bar J^3$ of $SO(2)_A$.\footnote{We consider  $J_3$ instead of ${\cal J}_3$
since we focus on the purely bosonic deformations but obviously the same argument extends
to the fermion bilinear pieces.} Notice that, had we spread the branes in the A plane,
the purely bosonic deformations would have been proportional to
\be
J^+ \bar J^- \Phi_{j;j,-j}e^{-q(j+1)\phi}\ ,\qq  J^- \bar J^+ \Phi_{j;-j,j}e^{-q(j+1)\phi}\ ,
\qq J^3 \bar J^3e^{-q(j+1)\phi} \ .
\ee
These are now invariant under $J^3+\bar J^3$, i.e.~the generator
of $SO(2)_B$, as they should.

\no
If we put now the NS5-branes on a regular polygon on the B plane,
which approaches a smooth circle in the $k \rightarrow \infty$, the symmetry
we expect is $SO(2)_A \times \mathbb{Z}_k$, while its continuous
limit should be $SO(2)_A \times SO(2)_B$. The commutator of the
$SO(2)_B$ generator $J^3+\bar J^3$ with the $j=m=\bar m=0$ deforming operator is zero, while
with the only other operator that is turned on, i.e.~that with $j=(k-2)/2$,
it yields
\be k J^+ \bar J^+ \Phi_{\frac{k-2}{2};\frac{k-2}{2},\frac{k-2}{2}} e^{-q (j+1)\phi}\
\ee
and similarly for its conjugate. In other words the deformation has charge
$k$ and therefore  is invariant under rotations by $2\pi/k$, i.e.~there is
indeed a discrete $\mathbb{Z}_k$ symmetry. More generally,
operators with a given $j$ have an invariance under $\mathbb{Z}_{2j+2}$
since they correspond to the deformations of \cite{Fotopoulos:2007rm} that
have indeed such a geometric symmetry.

\no
Before discussing the case of the bar, let us point out that there are
two relevant $SO(3)$ subgroups of $SU(2)_L \times SU(2)_R$, the latter being generated
by $J^\pm, J^3$ and $\bar J^\pm, \bar J^3$. The first, which we call
$SO(3)_B$ is generated by
\begin{equation}
SO(3)_B: \quad
J^3+\bar J^3,\quad J^++\epsilon \bar J^+, \quad J^-+\epsilon^{-1}\bar J^-\label{so3b}
\end{equation}
with $\epsilon$ an arbitrary complex number
and one of its quadratic invariants is
\begin{equation}
\epsilon^{-1} J^+ \bar J^- + \epsilon J^-\bar J^+ +2  J^3 \bar J^3\ .\label{so3binv}
\end{equation}
This invariant is part of the full Casimir constructed out of the generators
(\ref{so3b}); the latter is actually the sum of (\ref{so3binv}) and of the usual
Casimirs made out of
$J^3, J^\pm$ and $\bar J^3, \bar J^\pm$, for $SU(2)_{L}$ and $SU(2)_{R}$, respectively.

\no
The other relevant $SO(3)$
subgroup of $SU(2)_L \times SU(2)_R$, which will be denoted by $SO(3)_A$, is generated by
\begin{equation}
SO(3)_A:\quad
J^3-\bar J^3,\quad  J^++\epsilon \bar J^-, \quad J^-+
\epsilon^{-1}
\bar J^+
\end{equation}
and its invariant, analogous to  (\ref{so3binv}), is
\begin{equation}
\epsilon^{-1} J^+\bar J^+ +\epsilon J^-\bar J^- - 2  J^3 \bar J^3\ .
\end{equation}
Notice that demanding reality of these invariants requires $\epsilon$ to be a phase.
The corresponding $SO(2)_{A/B}$ subgroups are generated by $J^3\pm \bar J^3$
according to the convention we established at the beginning of this section.

\no
Let us consider now a configuration of NS5-branes arbitrarily spread along
a line in the B plane passing by the center. Such a line is described
by $B = e^{i \varphi} B^*$ where $\varphi$ is twice the angle it makes with the
$x^8$ axis. For $j=0$ the standard trace and the tilde one are identical and we
find that
the couplings $\lambda_{0;1,1}$,
$\lambda_{0;-1,-1}$ and $\lambda_{0;0,0}$ are related as
\be
\lambda_{0;-1,-1}= e^{ -i \varphi} \lambda_{0;0,0}\ ,\qq
\lambda_{0;1,1} = e^{i \varphi} \lambda_{0;0,0}\ .
\ee
Subsequently,
the  purely bosonic part of the corresponding deformation is given by the operator
\be
 \lambda_{0;0,0}
(e^{i \varphi} J^+\bar J^+  + e^{-i \varphi}J^-\bar J^- + 2  J^3 \bar J^3)e^{-q \phi}\ ,
\ee
where the factor of
2 in front of the last term appears because in (\ref{lagdef})
we are instructed to add the complex conjugate of every term.
  According to
the discussion above, this operator is indeed invariant
under $SO(3)_A$ for $\epsilon=-e^{-i \varphi}$.

\no
A similar situation would have arisen
if we had put the NS5-branes on a bar in the A plane, with $SO(3)_B$ being now
 the relevant symmetry
group. Notice also that this argument works 
in reverse. If we have a configuration with $SO(3)_A$ symmetry (and such
that ${\rm tr} (B^2) \neq 0$), the unique invariant that depends on both
holomorphic and antiholomorphic currents is
$\epsilon^{-1} J^+\bar J^+ +\epsilon J^-\bar J^- - 2  J^3 \bar J^3$
and therefore it
dictates the following relations between the couplings: $\lambda_{0;1,1} = -\epsilon^{-1}
\lambda_{0;0,0}$ and $\lambda_{0;-1,-1} = -\epsilon
\lambda_{0;0,0}$. In other words, ${\rm tr} (B^2) = -\epsilon^{-1} {\rm tr} (B B^*)$
and  ${\rm tr} \big((B^*) ^2\big) = -\epsilon {\rm tr} (B B^*)$, therefore implying
that $\epsilon$ is a phase and furthermore that\footnote{Proof: we would
like to show that if ${\rm tr} (B^2) = e^{ i \varphi} {\rm tr} (B B^*)$ then
$B =  e^{i \varphi} B^*$. Define $\Gamma=e^{-i\varphi/2} B$ so that
${\rm tr} (\Gamma^2) =  {\rm tr} (\Gamma \Gamma^*)$. $B$ and hence
$\Gamma$ are diagonal matrices and let  $r_i e^{i \theta_i}$ by the elements of the latter.
We can re-write the trace condition on $\Gamma$ as
$\sum_i (r^i)^2 (e^{2 i \theta_i} -1) =0$. The real part of this equation
gives $-2 \sum_i (r^i)^2 \sin^2\theta_i =0$, therefore implying that $\theta_i$ is an integer
multiple of $\pi$.
Hence $\Gamma=\Gamma^*$ and consequently $ B=e^{i\varphi} B^*$.}
 $B =  -\epsilon^{-1} B^*$.
Hence, only a configuration of NS5-branes along a line can have
$SO(3)_{A/B}$ symmetry.

\no
Since operators with $j>0$ are also
turned on, corresponding to the couplings $\widetilde {\rm tr} (B^{2j+2})$ which are generically non-vanishing,
one should also check that the associated  deformations are $SO(3)_A$ invariant.
The purely bosonic piece of the deformation corresponding to  $\widetilde {\rm tr} (B^{2j+2})$ is
$J^+ \bar J^+ \Phi_{j;j,j} e^{-q (j+1)\phi}$
and we have to include also its conjugate  $J^- \bar J^- \Phi_{j;-j,-j} e^{-q (j+1)\phi}$.
These operators are separately  invariant under $J^3-\bar J^3$ as we have already pointed out. However, they are not invariant under the other two generators
of $SO(3)_A$ since they correspond to deformations of higher order in the deforming
parameter and we expect that the actual generators of the $SO(3)$ symmetry
are also corrected beyond the leading order. It would quite interesting 
to uncover the corrected form of the symmetry generators beyond the leading
order. However,  this task is tantamount to constructing the CFT underlying this
configuration and therefore it should be quite non-trivial.

\no
We have summarized the symmetries of various
NS5-brane configurations in fig.~1.

\vskip 0 cm
\begin{figure}[htp]
\begin{center}
\includegraphics[height= 8 cm,angle=0]{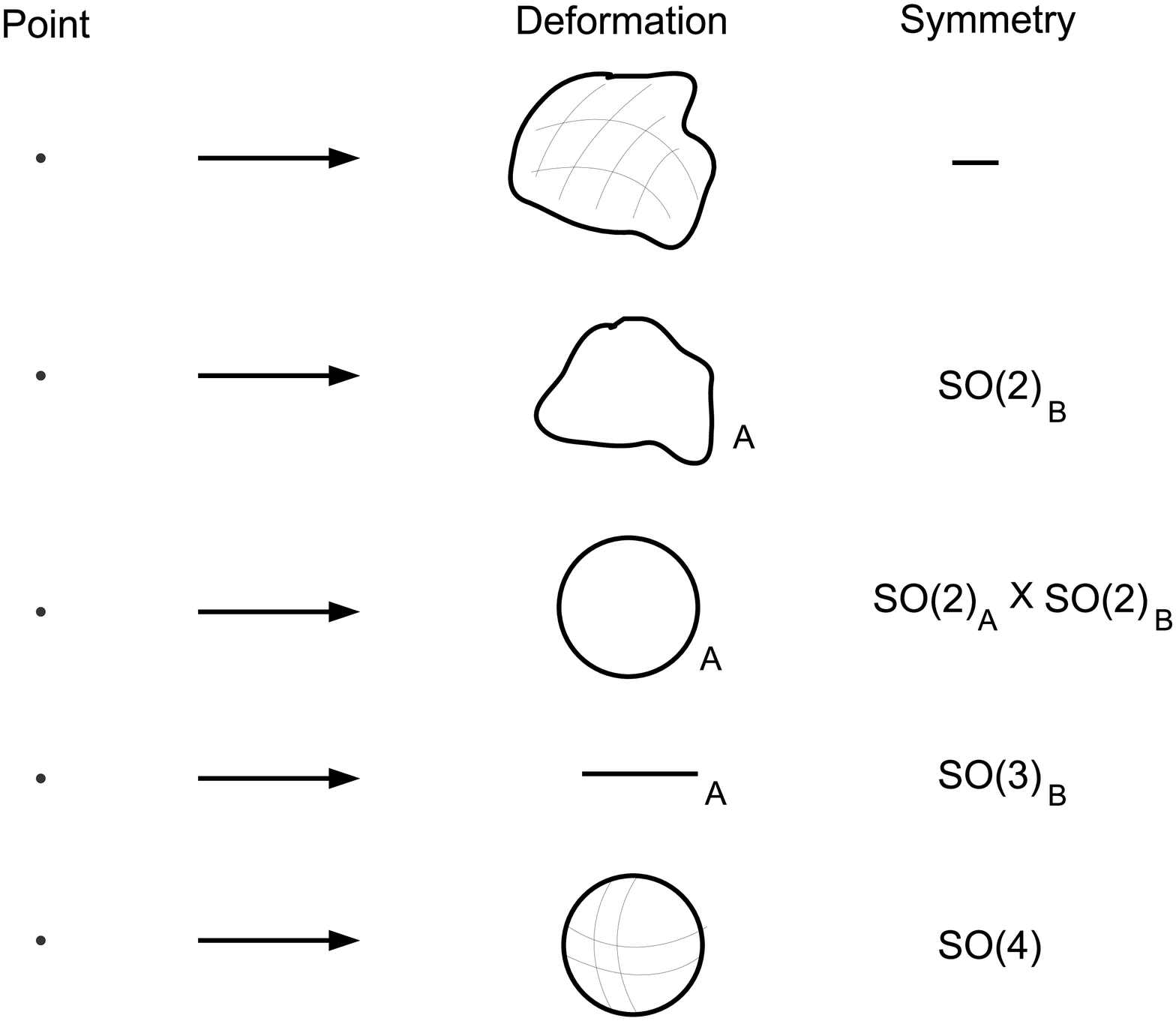}
\end{center}
\vskip -0.2 cm \caption{Various NS5-branes configurations and their symmetries in the
continuous ($k \rightarrow \infty$) limit: (i) generic distribution in $\mathbb{R}^4$ with
no symmetry, (ii) generic planar deformation with $SO(2)$ symmetry, (iii) circle with
$SO(2) \times SO(2)$ symmetry, (iv) bar with $SO(3)$ symmetry, (v) 3-sphere with
$SO(4)$ symmetry. }
\end{figure}

\section{\boldmath ``Non-holographic" CHS deformations
  \unboldmath}

A four-dimensional metric conformal to a hyperk\"ahler one
 supports generically ${\cal N}=(4,1)$ world-sheet supersymmetry
with torsion, provided that the conformal factor is a harmonic function of the
hyperperk\"ahler metric.
The background fields are
\be
d s^2 = H ds_{\rm HK}^2\ , \qq {\cal H}_{ijk}=\e_{ijk}{}^l \del_l H \ ,\qq e^{2\Phi} = H\ ,
\label{confhyp}
\ee
with $\nabla^2_{\rm HK} H=0$
and where the indices are raised with the hyperk\"ahler metric.
We present below one such example that was worked out as a gravity solution in
\cite{DeVaBo1, DeVaBo2} and we formulate it in the language of
$SU(2)_k \times \mathbb{R}_\phi$ deformations. The corresponding operator
is not chiral (or antichiral) primary and hence, according to our proposal in
subsection 2.5, does not correspond to a geometric NS5-brane deformation.

\no
Subsequently we examine
the case where the hyperperk\"ahler space is provided by $\mathbb{R}^4$
and present an interesting application concerning the leading correction of
$SU(2)_k \times \mathbb{R}_\phi$ towards the full NS5-brane solution. The correction
is triggered by one of the non-normalizable operators found in section 2
which preserve ${\cal N}=4$ supersymmetry but do not have holographic interpretation.

\subsection{Conformal Eguchi--Hanson metric}

This example
is provided by taking the Eguchi--Hanson as the hyperk\"ahler metric \cite{EguHan}
\be
{ds^2_{\rm EH}\ov 2 k}=  {r^4\ov r^4 -a^4 }\ dr^2 + r^2 (\s_1^2+\s^2_2) + {r^4 -a^4\ov r^2}\ \s_3^2 \ ,
\ee
where $\s_a$, $a=1,2,3$ are the Maurer--Cartan $SU(2)$ right-invariant 1-forms, normalized as
\be
d\s_a =  \e_{abc}\s_b \wedge \s_c\ .
\ee
We have also introduced an overall scale $2 k$ for later convenience.
Assuming that the conformal factor $H$ depends only on the radial variable $r$, we easily establish that
\be
H =  {A\ov 2 a^2} \ln \left(\frac{r^2+a^2}{r^2-a^2}\right)+ B \ ,
\label{hgfhee}
\ee
where $A$ and $B$ are integration constants.
In addition, we note that the antisymmetric NS 3-form
field strength is independent of $a$ and is given
by
\be
{\cal H}  = 2 \s_1 \wedge \s_2 \wedge \s_3\ .
\label{antti}
\ee
We select $A=1$ and $B=0$ so that for small $a$, or equivalently large $r$,
the conformal factor becomes
\be
H = {1\ov r^2}\left[1 + {a^4\ov 3 r^4} + {\cal O}\left(a^8\ov r^{8}\right)\right] \ .
\ee

\no
In this limit the background corresponds to a deformation of
$SU(2)_k\times \mathbb{R}_\phi$,
with the leading worldsheet correction being proportional to
\be
a^4 \left(J^1 \bar K^1 + J^2 \bar K^2 -2 J^3 \bar K^3\right)  e^{-2 q \phi}\ ,
\label{dfged}
\ee
where we have changed variables 
as
\be
r=e^{q\phi/2} \qq \Longrightarrow\qq \Phi = -{q\ov 2}\phi 
\label{jhgrr}
\ee
in order to make the dilaton linear to leading order.
We have also performed a
$\phi$-dependent reparametrization that keeps the coefficient of $d\phi^2 $ equal to one.

\no
The currents that appear in the deformation (\ref{dfged}) are given by
\be
J^a = -i {\rm tr}( \del g  g^{-1}\tau^a)\ ,\quad \bar K^a =-i {\rm tr}(
\bar \del g g^{-1} \tau^a)\ ,\qq a=1,2,3\ ,
\ee
with $\tau^a$ being the Pauli matrices.
Note that, whereas the current $J^a$ is the chirally conserved
current of $SU(2)$, obeying $\bar \del J^a = 0$, the current $\bar K^a $ is not its antiholomorphic
counterpart, i.e. $\bar \del K^a \neq 0$. However, one can write
\be
\bar K^a = C^a{}_{b} \bar J^b \ ,\qq C_{ab} = \ha {\rm tr}(\tau_a g \tau_b g^{-1})\ ,
\ee
where the currents
\be
\bar J^a = -i {\rm tr}(g^{-1} \bar \del g \tau^a)\ ,
\ee
are indeed antiholomorphic obeying $\del \bar J^a = 0$.
The matrix $C_{ab}$ is in the adjoint representation.

\no
 In order to make contact with the expressions
for the $SU(2)$ primaries of spin 1 consider the group element in the spin
1/2-representation parametrized as
\begin{equation}
 g = \begin{pmatrix}
 g_{++} & g_{+-}\\
 g_{-+} & g_{--}
  \end{pmatrix} \ ,
\ee
from which we compute that\footnote{
We use the basis $\s^\pm = \s_1 \mp i \s_2$, $\s_3$, where the indices are raised and lowered
with the metric $\eta_{33} = 2 \eta_{+-}= 2 \eta_{+-}=1$.
}
\ba
&& C^{\pm\pm}=-2 g_{\pm\mp}^2 \ ,\qq C^{\pm\mp}=2 g_{\pm\pm}^2\ ,
\nonumber\\
&& C^{3\pm}=\pm 2 g_{\mp\mp} g_{\pm\mp}\ ,\qq C^{\pm 3}=\mp 2 g_{\pm\pm} g_{\pm\mp} \ ,
\qq C^{33}= g_{+-} g_{-+} + g_{++} g_{--}\ . \label{candg}
\ea
On the other hand let us recall the transformation rules
\be
\d_{\pm} g_{\mp a}   = g_{\pm a} \ ,\qq \d_3 g_{\pm a } = \pm \ha g_{\pm a}\ ,\qq a=\pm\ ,
\ee
for the left $SU(2)$ action on the group element, a similar one for the right action
and the fact that for spin $j$ state with $m=\bar m = j $ is given by
\be
\Phi_{j;j,j} = g_{++}^{2 j } \ .
\ee
The other members of the representation are then obtained by acting with the above transfromation
rules. It is important to normalize the states generated in this way in a fashion compatible
with the OPEs \eqn{sdjop}. This is done if
 \be
\Phi_{j;m\pm 1,\bar m} =  {1\ov j\mp m}\ \d_\pm \Phi_{j;m,\bar m}\ ,
 \ee
for the left as well as for the right $SU(2)$ transformations.

\no
Using this method we easily find
\ba
&& \Phi_{1;\pm 1, \pm 1}=g_{\pm\pm}^2 \ ,\qq \Phi_{1;\pm 1, 0}= g_{\pm\pm} g_{\pm\mp}\ ,
\nonumber\\
&& \Phi_{1;\pm 1, \mp 1}= g_{\pm\mp}^2 \ ,\qq \Phi_{1;0 ,\pm 1}= g_{\pm\pm} g_{\mp\pm}\  ,
\qq \Phi_{1;0,0} = \ha (g_{++} g_{--} + g_{+-} g_{-+})\ .
\ea
Comparing now these expressions with (\ref{candg}) we can finally write
the elements of the matrix $C^{ab}$ in terms of the primaries $\Phi_{j;m,\bar m}$
for
$j=1$:
\ba
&& C^{\pm\pm}=-2 \Phi_{1;\pm 1,\mp 1}\ ,\qq C^{\pm\mp}=2 \Phi_{1;\pm 1,\pm 1}\ ,
\nonumber\\
&& C^{3\pm}=\pm 2  \Phi_{1;0,\mp 1 }\ ,\qq C^{\pm 3}=\mp 2  \Phi_{1;\pm 1,0 }\ ,
\qq C^{33}=2 \Phi_{1;0,0}\ .
\label{sdh1}
\ea
The currents $\bar K^a$ can be written in terms of the antiholomorphic currents
$\bar J^a$ as
\ba
\bar K^\pm & = &  \ha C^{ \pm + }\bar J^- + \ha C^{ \pm - }\bar J^+ + C^{ \pm 3  }\bar J^3\ ,
\nonumber\\
\bar K^3 & = &  \ha C^{ 3 +  }\bar J^- + \ha C^{ 3 -  }\bar J^+ + C^{3 3 }\bar J^3\ ,
\label{fh12}
\ea
and by using the relations \eqn{sdh1} we obtain the explicit representation
\ba
\bar K^+ & = & -\Phi_{1;1,-1} \bar J^- + \Phi_{1;1,1} \bar J^+ - 2 \Phi_{1;1,0} \bar J^3\ ,
\nonumber\\
\bar K^- & = & \Phi_{1;-1,-1} \bar J^- - \Phi_{1;-1,1} \bar J^+ + 2 \Phi_{1;-1,0} \bar J^3\ ,
\label{fh1rr}
\\
\bar K^3 & = & \Phi_{1;0,-1} \bar J^- - \Phi_{1;0,1} \bar J^+ + 2  \Phi_{1;0,0} \bar J^3\ .
\nonumber
\ea

\no
We can now rewrite the deformation \eqn{dfged} as
\be
a^4 \left(\ha J^+ \bar K^- + \ha J^- \bar K^+ -2 J^3 \bar K^3\right)  e^{-2 q \phi}\ ,
\label{dfge1}
\ee
and then, by  using  \eqn{fh1rr}, we can finally express the deformation as
\ba
&& a^4 \Bigg[
\left(-\ha \Phi_{1;1,-1} J^- +\ha \Phi_{1;-1,-1} J^+ -2 \Phi_{1;0,-1}J^3\right) \bar J^-
\nonumber\\
&& \phantom{x} +   \left(\ha \Phi_{1;1,1} J^- -\ha \Phi_{1;-1,1} J^+ +2 \Phi_{1;0,1}J^3\right) \bar J^+
\\
&&\phantom{x} +  \left(- \Phi_{1;1,0} J^- + \Phi_{1;-1,0} J^+ -4 \Phi_{1;0,0}J^3\right) \bar J^3
\Bigg]  e^{-2 q \phi}\ .
\nonumber
\ea
Comparing the first line above with \eqn{fjkb1} we find that they match
for $j=1$ and $m=0$.
From \eqn{fhhf1} we get that $\m_\pm =\pm 1/\sqrt{2}$ and $\m=-2$. Then \eqn{fjkb1} reproduces
the first line in the expression above.
Similarly, the other two lines in the above expressions match precisely, up to a multiplicative
factor, with \eqn{fjkb1} for the
same values of $j,m,\m_\pm $ and $\m$.
Therefore, the results of the first section imply that
the holomorphic pieces of the deformation (\ref{dfged}) preserve
${\cal N}=4$ supersymmetry. Instead, the antiholomorphic ones,
given by $\bar K^\pm$ and $\bar K^3$ in (\ref{fh1rr}), preserve neither ${\cal N}=2$
nor ${\cal N}=4$ supersymmetry. Hence, the total supersymmetry of
the background (\ref{confhyp}) is ${\cal N}=(4,1)$, in agreement with the analysis
of \cite{DeVaBo1, DeVaBo2}.

\no
We note that, by considering more general solutions than \eqn{hgfhee}
for the conformal factor $H$, we can construct perturbations corresponding to operators
with spin $j>1$.

\no
There is also an interesting interpretation of the background (\ref{confhyp})
in terms of NS5-branes. It is easy to see that the backreaction of a configuration
of NS5-branes put transversely on a  hyperk\"ahler space corresponds
exactly to the fields in  (\ref{confhyp}). In the particular case where
the transverse hyperk\"ahler space is the orbifold limit of an ALE space, such
systems were studied
in the context of LST holography in \cite{Diaconescu:1998pj}.
The fact that worldsheet supersymmetry
is reduced from ${\cal N}=(4,4)$ to ${\cal N}=(4,1)$ can be understood from this point of view as follows. Consider type IIB theory where the worldvolume of a set of
parallel NS5-branes, with transverse $\mathbb{R}^4$, supports ${\cal N}_6=(1,1)$ supersymmetry (by
${\cal N}_6$ we mean six-dimensional supersymmetries). If instead of the NS5-brane
geometry we had a
hyperk\"ahler space, the supersymmetry in the remaining six-dimensional Minkowski
space would be ${\cal N}_6=(2,0)$. Hence, superimposing the NS5-branes
with the hyperk\"ahler space, so that they share a common six-dimensional
Minkowski spacetime,
 leads to ${\cal N}_6=(1,0)$ supersymmetry. The latter necessitates
the presence of ${\cal N}=(4,1)$ in the worldsheet theory, in accordance with
the previous discussion.
Had we considered
type IIA string theory, the NS5-brane supersymmetry would have been ${\cal N}_6=(2,0)$
whereas the hyperk\"ahler space would have preserved ${\cal N}_6=(1,1)$, therefore
leading to the same result.

\subsection{Conformal $\mathbb{R}^4$: beyond the near-horizon}

In the case where the hyperk\"ahler space in (\ref{confhyp}) is
$\mathbb{R}^4$, we deal with a background that corresponds to a configuration
of NS5-branes whose distribution is specified by the harmonic function $H$.
Such backgrounds generically  exhibit ${\cal N}=(4,4)$ superconformal invariance.
\no
A very simple but quite interesting application of that construction is the following.
Recall that the near-horizon geometry $SU(2)_k \times
\mathbb{R}_\phi$ arises from the original solution corresponding to a
 point-like configuration
of NS5-branes by going to the near-horizon limit $r \rightarrow \infty$, which
is tantamount to dropping the "1" from the harmonic function. We would like
to consider the restoration of the full solution, i.e.~creating a constant term
in the harmonic function, as a deformation of the CHS background.
This deformation reads
\be
 \left(J^1 \bar K^1 + J^2 \bar K^2 + J^3 \bar K^3\right)  e^{ q \phi}
 =  \left(\ha J^+ \bar K^- + \ha J^- \bar K^+ + J^3 \bar K^3\right)  e^{ q \phi}\ ,
\label{fullsoldef}
\ee
where we performed the usual coordinate redefinition $r=e^{q\phi/2} $.

\no
Using the explicit form (\ref{fh1rr}) of the currents $\bar K^3, \bar K^\pm$ we can
re-write this deformation as
\ba
&&  \Bigg[
\left(-\ha \Phi_{1;1,-1} J^- +\ha \Phi_{1;-1,-1} J^+ + \Phi_{1;0,-1}J^3\right) \bar J^-
\nonumber\\
&& \phantom{x} +   \left(\ha \Phi_{1;1,1} J^- -\ha \Phi_{1;-1,1} J^+ - \Phi_{1;0,1}J^3\right) \bar J^+\label{firstexp}
\\
&&\phantom{x} +  \left(- \Phi_{1;1,0} J^- + \Phi_{1;-1,0} J^+ +2 \Phi_{1;0,0}J^3\right) \bar J^3
\Bigg]  e^{ q \phi}\ .
\nonumber
\ea
Another form, equal to the previous, is obtained by using the fact that the $SU(2)$ part in
(\ref{fullsoldef}) is the quadratic Casimir and hence we can replace the
right-invariant currents $ J^a, \bar K^a$ with left-invariant ones $\bar J^a,  K^a$:
\begin{equation}
\left(\ha J^+ \bar K^- + \ha J^- \bar K^+ + J^3 \bar K^3\right)  e^{ q \phi}
=\left(\ha K^+ \bar J^- + \ha K^-  \bar J^+ + K^3 \bar J^3\right)  e^{ q \phi} \ .
\end{equation}
Hence, the deformation can be written in an equivalent way as
\ba
&&  \Bigg[
\left(-\ha \Phi_{1;-1,1} \bar J^- +\ha \Phi_{1;-1,-1} \bar J^+ + \Phi_{1;-1;0} \bar J^3\right)  J^-
\nonumber\\
&& \phantom{x} +   \left(\ha \Phi_{1;1,1} \bar J^- -\ha \Phi_{1;1,-1} \bar J^+ - \Phi_{1;1,0}\bar J^3\right)  J^+\label{secondexp}
\\
&&\phantom{x} +  \left(- \Phi_{1;0,1} \bar J^- + \Phi_{1;0,-1}\bar J^+ +2 \Phi_{1;0,0} \bar J^3\right) J^3
\Bigg]  e^{ q \phi}\ .
\nonumber
\ea

\no
In the first expression (\ref{firstexp}) we see that the ${\cal N}=4$ preserving
non-normalizable operator of section 2 appears for the values of $j=1, m=0$.
In the second expression (\ref{secondexp}) the same operator appears, this time in
in the antiholomorphic sector, with $j=1, \bar m=0$. Therefore the deformation
preserves ${\cal N}=(4,4)$ superconformal invariance as well as spacetime supersymmetry,
in accordance with the fact that the full NS5-brane solution
exhibits that amount of supersymmetry. Notice that
this deformation was expected to be non-normalizable since
the near-horizon CHS background and the full NS5-brane solutions have
different asymptotic geometries. For other values of $j, m, \bar m$ they should
correspond to more general solutions for the harmonic function $H$.

\vskip .4 in
\centerline{ \bf Acknowledgments}

\no 
We would like to thank A.~Fotopoulos and V.~Niarchos for helpful discussions
 and T.~Suyama for correspondence.
 The author K. Sfetsos acknowledges partial support provided
through the European Community's program ``Constituents,
Fundamental Forces and Symmetries of the Universe'' with contract
MRTN-CT-2004-005104.

\end{document}